\documentclass[fleqn,usenatbib,useAMS]{mn2e}
\usepackage{times} 
\usepackage{latexsym,amssymb,amsmath,amsfonts}
\usepackage{rotfloat} 
\usepackage{rotating} 
\usepackage{float} 
\usepackage{graphicx}
\usepackage{longtable}
\usepackage{url}
\usepackage{dcolumn}
\usepackage{placeins}
\usepackage{multirow} 
\usepackage{indentfirst}
\usepackage{subfloat}
\usepackage{rotating} 
\usepackage[breaklinks,colorlinks,citecolor=blue,linkcolor=magenta]{hyperref}  
\begin{document}

\newcommand{\sqcm}{cm$^{-2}$}  
\newcommand{\lya}{Ly$\alpha$}
\newcommand{\lyb}{Ly$\beta$}
\newcommand{\lyg}{Ly$\gamma$}
\newcommand{\lyd}{Ly$\delta$}
\newcommand{\hi}{\mbox{\tiny H\,{\sc i}}}
\newcommand{\HI}{\mbox{H\,{\sc i}}}
\newcommand{\HII}{\mbox{H\,{\sc ii}}}  
\newcommand{\Htot}{\mbox{{\rm H}}}	
\newcommand{\HeI}{\mbox{He\,{\sc i}}}
\newcommand{\HeII}{\mbox{He\,{\sc ii}}}
\newcommand{\HeIII}{\mbox{He\,{\sc iii}}}  
\newcommand{\OI}{\mbox{O\,{\sc i}}}
\newcommand{\OII}{\mbox{O\,{\sc ii}}}
\newcommand{\OIII}{\mbox{O\,{\sc iii}}}
\newcommand{\OIV}{\mbox{O\,{\sc iv}}}
\newcommand{\OV}{\mbox{O\,{\sc v}}}
\newcommand{\OVI}{\mbox{O\,{\sc vi}}}
\newcommand{\OVII}{\mbox{O\,{\sc vii}}}
\newcommand{\OVIII}{\mbox{O\,{\sc viii}}} 
\newcommand{\NaI}{\mbox{N\,{\sc i}}}
\newcommand{\CI}{\mbox{C\,{\sc i}}}
\newcommand{\CII}{\mbox{C\,{\sc ii}}}
\newcommand{\CIII}{\mbox{C\,{\sc iii}}}
\newcommand{\CIV}{\mbox{C\,{\sc iv}}}
\newcommand{\CIVdblt}{\CIV~$\lambda\lambda$1548, 1550}
\newcommand{\CV}{\mbox{C\,{\sc v}}}
\newcommand{\CVI}{\mbox{C\,{\sc vi}}}  
\newcommand{\SiII}{\mbox{Si\,{\sc ii}}}
\newcommand{\SiIII}{\mbox{Si\,{\sc iii}}}
\newcommand{\SiIV}{\mbox{Si\,{\sc iv}}}
\newcommand{\SiXII}{\mbox{Si\,{\sc xii}}}   
\newcommand{\SII}{\mbox{S\,{\sc ii}}}
\newcommand{\SIII}{\mbox{S\,{\sc iii}}}
\newcommand{\SIV}{\mbox{S\,{\sc iv}}}
\newcommand{\SV}{\mbox{S\,{\sc v}}}
\newcommand{\SVI}{\mbox{S\,{\sc vi}}}  
\newcommand{\NI}{\mbox{N\,{\sc i}}}   
\newcommand{\NII}{\mbox{N\,{\sc ii}}}   
\newcommand{\NIII}{\mbox{N\,{\sc iii}}}     
\newcommand{\NIV}{\mbox{N\,{\sc iv}}}   
\newcommand{\NV}{\mbox{N\,{\sc v}}}    
\newcommand{\PV}{\mbox{P\,{\sc v}}} 
\newcommand{\NeIV}{\mbox{Ne\,{\sc iv}}}   
\newcommand{\NeV}{\mbox{Ne\,{\sc v}}}   
\newcommand{\NeVI}{\mbox{Ne\,{\sc vi}}}   
\newcommand{\NeVII}{\mbox{Ne\,{\sc vii}}}   
\newcommand{\NeVIII}{\mbox{Ne\,{\sc viii}}}   
\newcommand{\NeIX}{\mbox{Ne\,{\sc ix}}}   
\newcommand{\NeX}{\mbox{Ne\,{\sc x}}} 
\newcommand{\MgI}{\mbox{Mg\,{\sc i}}}
\newcommand{\MgII}{\mbox{Mg\,{\sc ii}}}  
\newcommand{\MgX}{\mbox{Mg\,{\sc x}}}   
\newcommand{\FeII}{\mbox{Fe\,{\sc ii}}}  
\newcommand{\FeIII}{\mbox{Fe\,{\sc iii}}}   
\newcommand{\NaIX}{\mbox{Na\,{\sc ix}}}   
\newcommand{\ArVIII}{\mbox{Ar\,{\sc viii}}}   
\newcommand{\AlXI}{\mbox{Al\,{\sc xi}}}   
\newcommand{\CaII}{\mbox{Ca\,{\sc ii}}}  
\newcommand{\zabs}{$z_{\rm abs}$}
\newcommand{\zmin}{$z_{\rm min}$}
\newcommand{\zmax}{$z_{\rm max}$}
\newcommand{\zqso}{$z_{\rm QSO}$}
\newcommand{\zgal}{$z_{\rm gal}$}
\newcommand{\degree}{\ensuremath{^\circ}}
%...............................................................................
\newcommand{\lapp}{\mbox{\raisebox{-0.3em}{$\stackrel{\textstyle <}{\sim}$}}}
\newcommand{\gapp}{\mbox{\raisebox{-0.3em}{$\stackrel{\textstyle >}{\sim}$}}}
\newcommand{\be}{\begin{equation}}
\newcommand{\en}{\end{equation}}
\newcommand{\di}{\displaystyle}
\def\tworule{\noalign{\medskip\hrule\smallskip\hrule\medskip}} %double rule.%
\def\onerule{\noalign{\medskip\hrule\medskip}} %single rule.%
\def\bl{\par\vskip 12pt\noindent}
\def\bll{\par\vskip 24pt\noindent}
\def\blll{\par\vskip 36pt\noindent}
\def\rot{\mathop{\rm rot}\nolimits}
\def\alf{$\alpha$}
\def\lam{$\lambda$}
\def\refff{\leftskip20pt\parindent-20pt\parskip4pt}
\def\kms{km~s$^{-1}$}
\def\zem{$z_{\rm em}$} 
\def\cc{$\rm cm^{-3}$}
\def\vrel{$v_{\rm rel}$}
\def\cmsq{cm$^{-2}$}
\def\cmcb{cm$^{-3}$}
\def\etal{et~al.\ }
\newcommand{\CLOUDY}{\mbox{\scriptsize{CLOUDY}}}
%---------------------------------------------------------------------------------------------------------------------------

\title[Revisiting the \zabs\ = 0.93 system towards PG~1206+459]{Understanding the strong intervening \OVI\ absorber at $z_{\rm abs} \sim 0.93$ towards PG1206+459 \thanks{Based on observations made with the NASA/ESA {\sl Hubble Space Telescope}, obtained from the data archive at the Space Telescope Science Institute, which is operated by the Association of Universities for Research in Astronomy, Inc., under NASA contract NAS 5-26555.}}   
\author[Rosenwasser et al.]
{
\parbox{\textwidth}{ 
B. Rosenwasser$^{1,2}$,
S. Muzahid$^{1,3}$,  
J. C. Charlton$^{1}$, 
G. G. Kacprzak$^{4}$, 
B. P. Wakker$^{2}$, and          
C. W. Churchill$^{5}$    
}
\vspace*{10pt}\\ 
$^{1}$The Pennsylvania State University, 525 Davey Lab, University Park, State College, PA 16802, USA \\   
$^{2}$Department of Astronomy, University of Wisconsin, Madison, WI 53706, USA  \\  
$^{3}$Leiden Observatory, Leiden University, PO Box 9513, 2300 RA Leiden, the Netherlands \\ 
$^{4}$Swinburne University of Technology, Victoria 3122, Australia  \\ 
$^{5}$New Mexico State University, Las Cruces, NM 88003, USA  
}   
\date{Accepted to MNRAS}
\maketitle
\label{firstpage}
%---------------------------------------------------------------------------------------------------------------------------

%============================== ABSTRACT =================================
\begin{abstract} 
We have obtained new observations of the partial Lyman limit absorber at \zabs$=0.93$ towards quasar PG~1206+459, and revisit its chemical and physical conditions. The absorber, with $N(\HI)\sim10^{17.0}$~\sqcm\ and absorption lines spread over $\gtrsim$1000~\kms\ in velocity, is one of the strongest known \OVI\ absorbers at $\log N({\OVI})=$~15.54$\pm$0.17. Our analysis makes use of the previously known low-(e.g. \MgII), intermediate-(e.g. \SiIV), and high-ionization (e.g., \CIV, \NV, \NeVIII) metal lines along with new $HST/$COS observations that cover \OVI, and an $HST/$ACS image of the quasar field. Consistent with previous studies, we find that the absorber has a multiphase structure. The low-ionization phase arises from gas with a density of $\log (n_{\rm H}/\rm cm^{-3})\sim-2.5$ and a solar to super-solar metallicity. The high-ionization phase stems from gas with a significantly lower density, i.e. $\log (n_{\rm H}/\rm cm^{-3}) \sim-3.8$, and a near-solar to solar metallicity. The high-ionization phase accounts for all of the absorption seen in \CIV, \NV, and \OVI. We find the the detected \NeVIII, reported by \cite{Tripp2011}, is best explained as originating in a stand-alone collisionally ionized phase at $T\sim10^{5.85}~\rm K$, except in one component in which both \OVI\ and \NeVIII\ can be produced via photoionization. We demonstrate that such strong \OVI\ absorption can easily arise from photoionization at $z\gtrsim1$, but that, due to the decreasing extragalactic UV background radiation, only collisional ionization can produce large \OVI\ features at $z\sim0$. The azimuthal angle of $\sim88$\degree\ of the disk of the nearest ($\rm 68~kpc$) luminous ($1.3L_*$) galaxy at $z_{\rm gal}=0.9289$, which shows signatures of recent merger, suggests that the bulk of the absorption arises from metal enriched outflows. 
\end{abstract}     
%=========================== KEY WORDS =================================== 
% %=========================== INTRODUCTION ==============================   
\section{Introduction} 
\label{sec_intro}  

Baryons reside in both the luminous central regions of galaxy halos and the diffuse circumgalactic medium (CGM) seen primarily in absorption. The accretion and feedback processes involved in galaxy evolution extend into the CGM, where spectral absorption line diagnostics can constrain the column densities, kinematics, ionization conditions, and metallicity of the absorbing gas. Circumgalactic gas is a fundamental component of galaxies, together with the interstellar medium (ISM), stars, and dark matter halo, and a complete picture of galaxy evolution should explain its observed properties and its connection with the host-galaxies at different cosmic epochs.

Numerical simulations predict that ``cold'' accretion of $T\sim 10^4 - 10^5$~K gas can penetrate the halos of galaxies still forming stars, with halo masses $< 10^{12} M_\odot$, while more massive halos shock heat and maintain the accreting gas at higher ($\sim 10^6$~K) temperatures (e.g.,\cite{Keres2005, Dekel2006, KeresHern2009}). Simulations also require a prescription for some form of large scale galactic feedback, both stellar \citep[e.g.,][]{Vielleux2005} and active galactic nuclei (AGN), in order to avoid overproduction of stars and to enrich the CGM and the intergalactic medium \citep[IGM, e.g.,][]{Keres2009,Dave2011b,Dave2011a}. These two processes, inflows and outflows, are the main components of current simulations and require detailed constraints provided by observational studies of the CGM.

The existence of galactic scale outflows is well-established for local galaxies with star-formation rate densities above $0.1 \text{ }M_\odot~\rm yr^{-1 }~kpc^{-2}$ \citep{Heckman2002}. At higher redshifts, where this threshold value is more commonly achieved, outflows are observed to be ubiquitous, for example at $z \sim 1.5$  \citep[]{Rupke2005,Weiner2009,Rubin2014,Benzhu2015} and $z\sim3$ \citep[]{Pettini2001,Shapley2003}. The usual tracers for these outflows are neutral or singly ionized species (e.g., \NaI, \MgI, and \MgII) that stem from material that has been entrained by supernovae and/or stellar winds. Higher ionization tracers of winds (e.g., \OVI, \NeVIII) in these ``down-the-barrel'' absorption lines studies are hard to detect since these lines lie in the far- and extreme-ultraviolet (FUV, EUV) region of the spectrum, where the continuum from the galaxy is usually faint.    

\citet{Grimes2009} carried out a study of local starbursts in the FUV using the Far Ultraviolet Spectroscopic Explorer (FUSE). They detect \OVI\ in nearly all of their 16-galaxy sample with column densities $15.3>\log N(\OVI)>14.0$ and outflow velocities of the highly ionized gas up to $\sim$300~\kms. They confirm previous findings that the star formation rate (SFR) and specific SFR (sSFR) of the host galaxy is positively correlated with the outflow velocity. The \OVI\ in their study extends to higher velocities than the neutral and photoionized gas, which they interpret as arising in a cooling, hot gas flow seen in X-ray. \cite{Weiner2009} also report a dependence on galaxy mass and color with outflow velocity and equivalent width, though substantial outflows are still observed for the low-mass, low-SFR galaxies in their sample. 

The galactic winds characteristic of low and high mass galaxies are driven by the mechanical energy supplied by supernovae and winds from massive stars. These winds generate an expanding shell, which fragments due to Raleigh-Taylor instabilities, allowing for the hot wind fluid to expand into the halo as bipolar outflows \citep{Heckman2002}. Large amounts of dense interstellar gas (references above) can escape into the halo with this hot wind fluid. The fate of these winds as they enter the CGM is largely unknown and require a sufficiently bright background UV continuum source that can probe the intervening outflow.

There has been much effort to characterize the gas in the CGM since the installation of the Cosmic Origins Spectrograph on $HST$ \citep{Green2012}. These studies have focused on gas tracing individual outflows \citep{Tripp2011,Muzahid2014,Muzahid2015} as well as global properties of the CGM presumably enriched via outflows \citep{Tumlinson2011,Bordoloi2014,Kacprzak15}. \citet{Tumlinson2011} show that the highly ionized transition \OVI, with $\log N(\OVI)> 14.3$, is preferentially detected around $L^*$ star-forming galaxies, whereas lower ionization transitions, e.g. \MgII, have high covering fractions around both star-forming and passive galaxies \citep{Thom2012,Werk2013}. The mass in metals and hydrogen in the CGM of $L^*$ galaxies can be substantially larger than that found in stars and the ISM and may resolve the galactic missing baryons problem \citep{Werk2014,Peeples2014,Prochaska2017}.

The \OVI~$\lambda\lambda$1031,1037 doublet is particularly important in the search for the missing baryons because its high abundance and ionization potential allow it to trace a range of physical environments. In the 54 systems with \OVI\ and \HI\ studied by \cite{Savage2014} with $13.1 < \log N(\OVI) <14.8$, 69\% traced cool $\sim10^4$~K photoionized gas while 31\% traced warm $\sim 10^5-10^6$~K gas. 40 out of the 54 \OVI\ systems have associated galaxies within 1~$\rm Mpc$, most within 600~$\rm kpc$, which are higher impact parameters than those probed by \citet{Tumlinson2011}. Intergalactic warm \OVI\ absorbers constitute the warm hot intergalactic medium (WHIM) that is thought to contain many of the cosmological missing baryons.

A particularly interesting absorption line system is the Lyman limit system (LLS) towards PG~1206+459 at \zabs~$\sim0.93$, with $N(\HI)\sim10^{17.0}$~\sqcm\ . This system has strong low and high ionization absorption lines, including the strongest known \OVI\ absorption of any intervening absorber, and spans a large ($\sim$1500 \kms) velocity range. There has been three focused studies of this system so far \citep{Churchill1999,Ding2003a,Tripp2011}, and it was included in the \citet{Fox2013} study of z$<$1 LLSs. \citet{Churchill1999} first identified the strong \MgII\ system in a HIRES spectrum ($R\sim6$~\kms) and classified the three apparent sub-systems at \zabs\ = 0.9254, 0.9276, and 0.9243 as systems A, B, and C, respectively.  The initial study of the high ionization transitions \CIV, \NV, and \OVI\ was limited by the low resolution Faint Object Spectrograph (FOS) spectrum. Based on the large velocity spread and slight overdensity of galaxies in the quasar field, they entertain the idea of the absorption arising in a group environment.

The study of the complex continued by \citet[hereafter D03]{Ding2003a} with an $R=15$~\kms\ E230M Space Telescope Imaging Spectrograph (STIS) spectrum with coverage of \lya, \SiII, \CII, \SiIII, \SiIV, \CIV\ and \NV. The authors favored a two-phase photoionization model where \SiIV\ traces the same gas as \MgII, and \CIV\ and \NV\ trace a second phase. The high ionization phase could also account for the equivalent width of \OVI\ seen in the low-resolution FOS spectrum. The \lya\ in the STIS spectrum and Lyman series covered by FOS placed constraints on the metallicity of the gas to be solar or super-solar. D03 also presented a WIYN $i$-band image of the quasar field and CryoCam spectra of the candidate galaxies. They detected an [\OII]~$\lambda$3727 emission line from one galaxy at $z=0.9289\pm0.0005$, placing it $\sim+200$~\kms\ relative to System B. They also report a marginal detection of another galaxy (G3 in their image) in a Fabry-Perot image tuned to redshifted [\OII] at $z = 0.93$.

The first medium resolution FUV spectrum of the absorber was obtained with the G130M and G160M gratings on the Cosmic Origins Spectrograph (COS) by \citet[hereafter T11]{Tripp2011}, with coverage of the Lyman break and many transitions blue-ward of 912~\AA. The authors also presented an MMT spectrum of the associated galaxy and classified it as a post-starburst galaxy based on Balmer absorption and [\OII], [\NeV] emission lines. Using the detected \NeVIII~$\lambda\lambda$770,780 doublet, they favor a collisional ionization model of the gas producing \NV\ and \NeVIII. Considering the large metallicities in the different components and the galaxy properties, they attribute the strong metal absorption to a large scale galactic outflow.

In the this paper we will present the COS/G185M spectrum of PG~1206+459 with coverage of the \OVI\ doublet and an $HST$ image of the associated galaxy. Section~\ref{sec:obs} details the observations and data analysis procedure. In Section~\ref{sec:PImod} we present photoionization models of each absorption system. In Section~\ref{sec:CI} we discuss collisional ionization models.  The galaxies that are detected near the quasar sightline are presented in Section~\ref{sec:galaxy}. We discuss our results in Section~\ref{sec:discussion} followed by conclusions in Section~\ref{sec:conclusions}. Throughout this paper we adopt an $H_{0}=70$~\kms~$\rm Mpc^{-1}$, $\Omega_{\rm M}=0.3$, and $\Omega_{\rm \Lambda}=0.7$ cosmology. Solar abundances of heavy elements are taken from \citet{Asplund2009}. All the distances given are proper (physical) distances.

% %=========================== OBSERVATIONS AND DATA REDUCTION ==============================   
\section{Observations and Data Reduction}
\label{sec:obs}

\subsection{Absorption Data}
\label{subsec:absdata}

A medium resolution ($R\sim$18,000), high signal-to-noise ratio ($S/N\sim$40 per resolution element) FUV spectrum of PG~1206+459 (\zem\ = 1.164) was obtained using $HST/$COS during observation Cycle-17 under program ID: 11741. These observations consist of G130M and G160M FUV grating exposures covering the wavelength range 1150--1800 \AA, and they were the basis of the study by T11. In order to add constraints from \OVI\ to the study, we obtained a NUV spectrum using COS/G185M grating with a similar resolution, covering 1775--1818 \AA, 1878--1921 \AA, and 1983--2025 \AA, and with a typical $S/N\sim 10$ per resolution element during Cycle-19 under program ID: 12466. The properties of COS and its in-flight operations can be found in \citet{Osterman2011} and \citet{Green2012}. The data were retrieved from the $HST$ archive and reduced using the STScI CALCOS v2.21 pipeline software. Individual exposures were aligned and coadded using the methods described in \citet{Hussain2015,Wakker15}.

% 
%==================================================================================
\begin{figure*}
\centerline{
\vbox{
\centerline{\hbox{
\includegraphics[width=0.65\textwidth,angle=00]{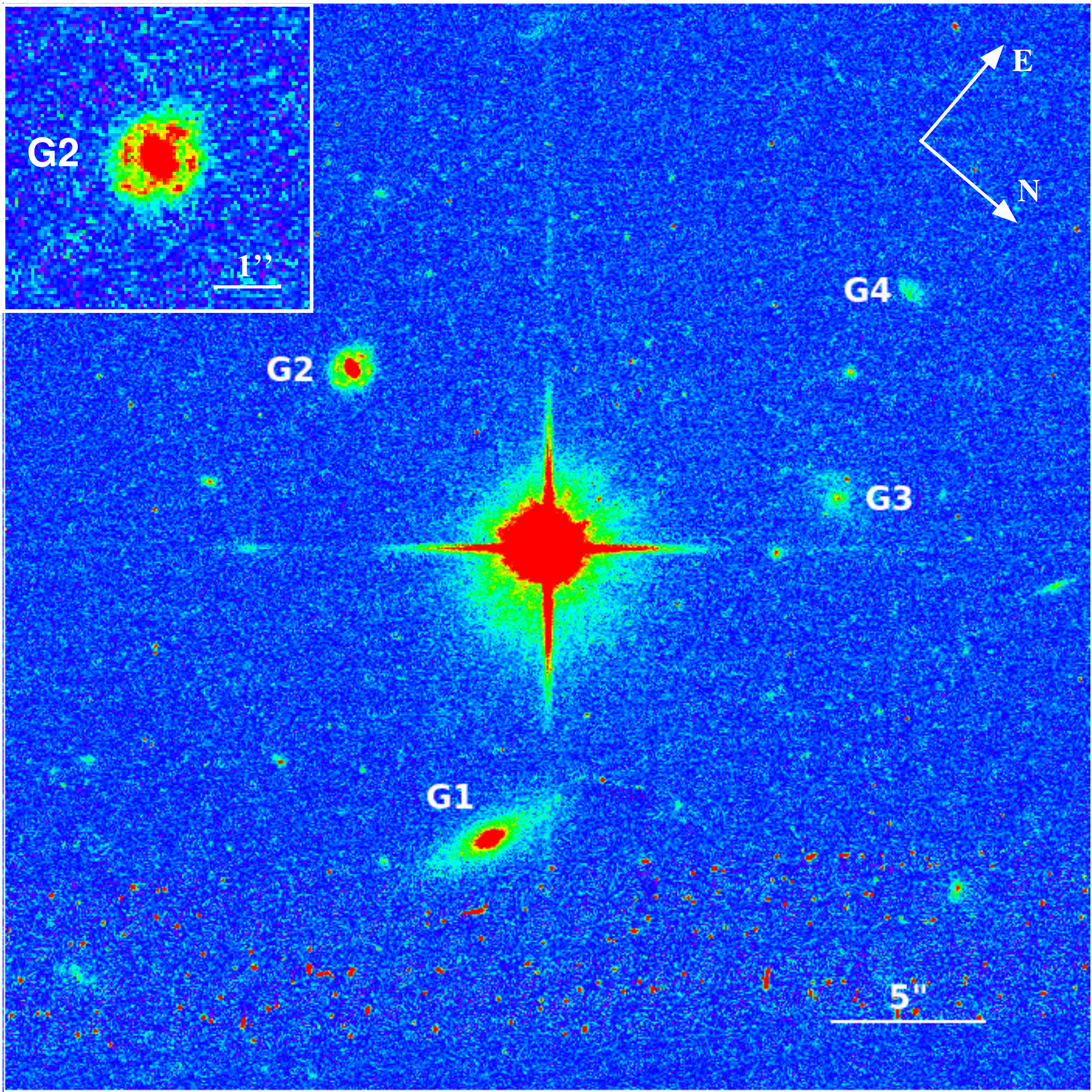}  
\includegraphics[width=0.35\textwidth,angle=00]{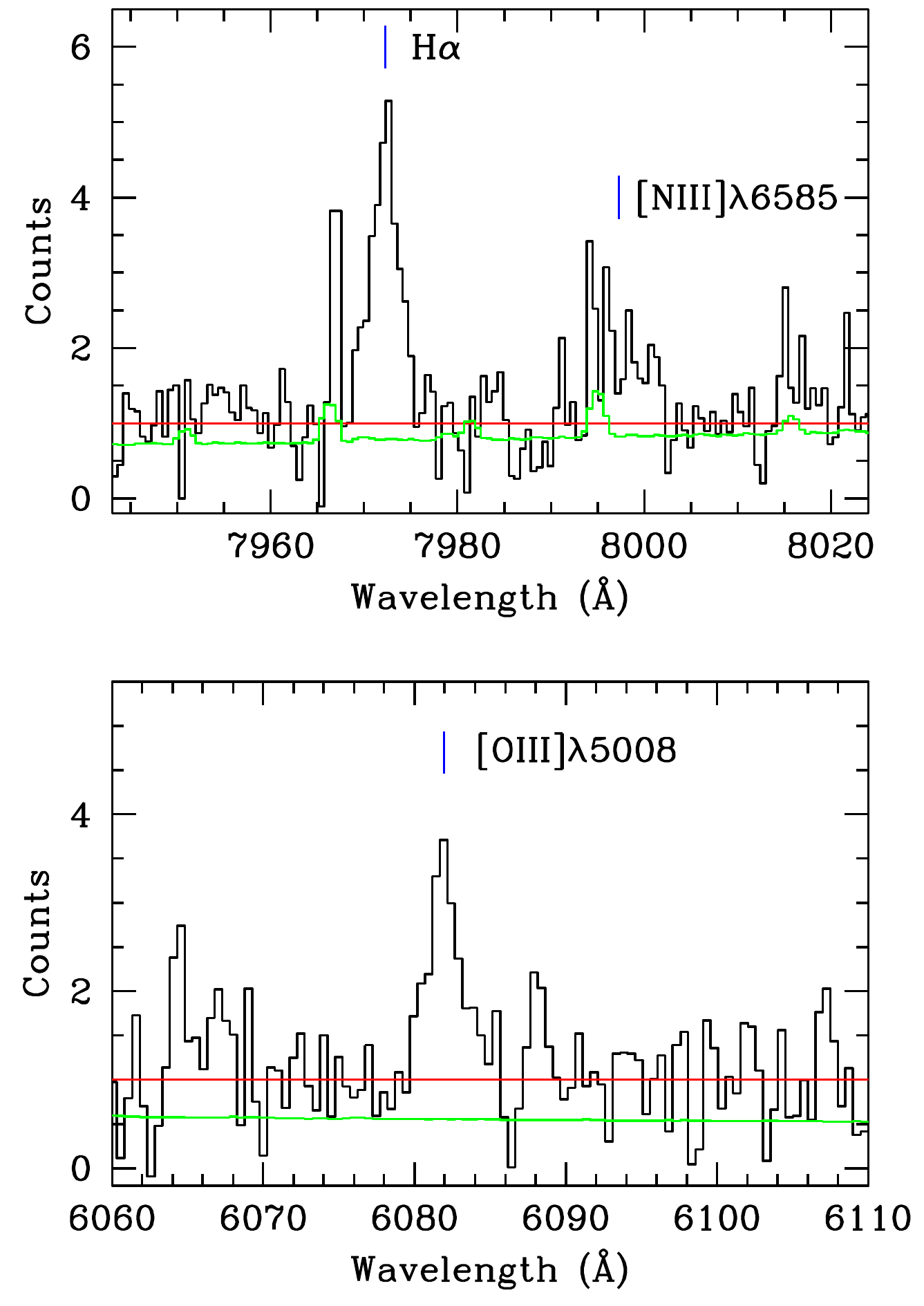} 
}}
}}
\caption{{\bf Left:} An $HST/$ACS F814W image of the PG~1206+459 field with galaxies G1--G4 labelled. The only galaxy which has spectroscopic redshift consistent with the \zabs\ is G2. A ring-like structure of G2 is evident from the zoomed in view shown in the inset. {\bf Right:} Selected regions from the Keck/ESI spectrum of galaxy G1 showing the emission lines of \OIII~$\lambda$5008, H$\alpha$ and a sky-line blended \NII~$\lambda$6585. The identified H$\alpha$ and \OIII~$\lambda$5008 lines determine a redshift of $z=0.2144\pm0.00002$. Therefore the galaxy does not contribute to the absorption complex we study here.}          
\label{fig:HSTimage}    
\end{figure*} 
%==================================================================================
%  

The reduced co-added spectra were binned by three pixels, as the COS FUV data, in general, are highly oversampled (i.e. six raw pixels per resolution element). All measurements and analysis presented in this article were performed on the binned data. NUV data with 2 raw pixels per resolution element, however, were not binned. Continuum normalization was done by fitting the line-free regions with smooth lower-order polynomials.

Critical constraints were also provided by a $R=30,000$ {\it HST}/STIS spectrum, using the E230M grating, which covers 2270--3120~\AA\ and thus the \CIV\ and \NV\ for the $z=0.927$ system.  Similarly, a previously published $R=45,000$ Keck$/$HIRES spectrum covers \MgII, \FeII, and \MgI. We refer the reader to D03 for further information about the observations and data reduction procedures for the STIS and HIRES spectra.

% 
%=====================================================================================
\begin{table}
\begin{center}
\caption{Galaxies in the PG~1206+459 field}
\begin{tabular}{ccccc}
\hline
Galaxy  & \zgal                &    $L_B$   &  $D$   &  $r_h$  \\ 
        &                      &    ($L^*$) & ($\rm kpc$) &  ($\rm kpc$)     \\   
\hline
G1      &   0.2144$\pm$0.00002$^a$ &    $0.03$   &   31  &      3.4    \\ 
G2      &   0.9289$\pm$0.0005$^b$  &     $1.3$   &   68  &      2.7    \\
G3      &   0.93$^c$           &    $0.44$   &   74  &      5.5    \\
G4      &   0.93$^c$           &    $0.12$   &  113  &      2.6    \\
\hline
\label{tab:gal}      
\end{tabular}
\raggedright{}
\end{center} 
\raggedright{
$^a$This work. $^b$From D03.  
$^c$Assuming a redshift of $z=0.93$ the other properties are listed.}  Column 1 is galaxy identification, column 2 is the redshift of the galaxy, column 3 is the B-band luminosity in units of L$^*$, column 4 is the impact parameter of the galaxy, and column 5 is the half-light radius of the galaxy
\end{table}
%=====================================================================================
%  

\subsection{Galaxy Data}
\label{subsec:galdata}  
An $HST/$ACS image of the PG~1206+459 field (with the F814W filter) was obtained as part of a public snapshot survey (PID: 13024) intended for studying galaxies associated with \OVI\ and/or \NeVIII\ absorbers. The exposure time for this snap observation was 8059 seconds. The image, displayed in Fig.~\ref{fig:HSTimage}, shows four galaxies within several arcseconds of the quasar sightline. These are the same four galaxies (G1--G4) that were detected within 6\arcsec\ of the quasar in a WIYN image of the PG 1206+459 field, published in Fig. 4 of D03. The magnitudes (in the Vega system) of the four galaxies were determined with 1.5$\sigma$ isophotes in Source Extractor \citep{Bertin1996}. GIM2D \citep{Simard2002} was utilized to find the inclination angles ($i$) and orientation angles ($\Phi$) of the galaxies as described in \cite{Kacprzak2011}. $\Phi =0$\degree\ corresponds with alignment of the quasar line of sight with the galaxy projected major axis, and $\Phi = 90$\degree\ with alignment of the quasar line of sight with the galaxy projected minor axis.

A spectrum of galaxy G1 was obtained using the Keck Echelle Spectrograph and Imager \citep[ESI;][]{Sheinis2002} on 2014 April 25 with an exposure time of 1000 seconds.  We used the 20$''$ long and 1$''$ wide slit and used 2$\times$2 on-chip CCD binning.  The ESI wavelength coverage is 4000$-$10,000~{\AA}, which provides coverage of all nebular optical emission lines for low-to-intermediate redshift galaxies with a velocity dispersion of 22~km~s$^{-1}$~pixel$^{-1}$ when binning by two in the spectral direction (FWHM$\sim$90~km~s$^{-1}$). The spectrum was reduced using the standard Echelle package in IRAF along with standard calibrations and was vacuum and heliocentric velocity corrected. The details of the galaxy properties are summarized in Table~\ref{tab:gal}  and are further discussed in Section~\ref{sec:galaxy}.

%=====================================================================================
\section{Photoionization Models}   
\label{sec:PImod}

In this section we present our methods and results from photoionization modeling of the absorption complex at \zabs$\sim$0.927 towards quasar PG~1206+459.

\subsection{Method for Modeling}

Our procedure for photoionization modeling is similar to that employed in previous studies as presented in \cite{Charlton2003,Zonak2004,Ding2003a,Ding2003,Ding2005,Masiero2005}. The goal of this approach is to minimize the number of gas phases while providing an adequate fit to the data. Although the resulting solution is not unique, it is the simplest plausible solution.  By ``phase'' of gas, we mean gas within a small range of temperature and density giving rise to absorption features with similar column densities, in this case across several absorption components.

We begin by taking the column density ($N$) and Doppler parameter ($b$) of each \MgII\ component from \cite{Churchill1999}, also used by D03, which were obtained by Voigt profile fitting using the program {\sc minfit} \citep[]{cwc-thesis}. The \MgII\ profiles are fit with relatively narrow and distinct components, given that the spectra are of high-resolution and high $S/N$,  thus this transition is the best starting point for optimizing the low-ionization phase model.

The low-ionization phase of the absorber is modeled for each of the \MgII\ components as a slab irradiated by the extragalactic UV background radiation (EBR) at $z=0.92$ as computed by \citet[][hereafter HM01]{Haardt2001}. The EBR is normalized at a hydrogen ionizing photon number density of $\log~[n_{\gamma}/\rm cm^{-3}] = -4.96$, appropriate for the given redshift. Since the exact shape and normalization of the EBR is uncertain, we consider the effect of using an alternative EBR model (and the host galaxy radiation field) in Section \ref{alterEBR}.

Photoionization models are run using the code {\sc cloudy} \citep[v13.03; last described by][]{Ferland2013}. For each of the Voigt profile components for \MgII, we run a series of {\sc cloudy} models using a grid of values for the ionization parameter, $U$ ($U=n_{\gamma}/n_{\rm H}$, where $n_{\rm H}$ is the total hydrogen number density), and the metallicity, $Z$, in units of the solar value. The abundance pattern could also be a free parameter, but for simplicity a solar pattern \citep[i.e.,][]{Asplund2009} is assumed unless otherwise noted. For each cloud, and for each point on the grid ($U$, $Z$), we iterate with different values of the total column density of hydrogen, $N_{\rm H}$, until a value of $N_{\rm H}$ is found for which the model slab reproduces the observed column density of {\MgII}.

At each grid point ($U$, $Z$) the {\sc cloudy} model also yields column densities for all other ionic transitions. With the temperature output from {\sc cloudy}, the thermal ($b_{th}$) and turbulent ($b_{nt}$) components of the \MgII\ $b$-parameter can be separated from the observed \MgII\ Doppler ($b$) parameter, $b(\MgII)$, using $b(\MgII)^2 = b_{th}^2 + b_{nt}^2$, where $b_{th}^2 = 2k_{\rm B}T/m_{\rm Mg}$. We then use the $b_{nt}$ to calculate the $b$-parameters for the other transitions. Using the $b$-parameters and the {\sc cloudy} model-predicted column densities, we generate a synthetic absorption spectrum convolved with the line spread function of the relevant spectrograph.  Many grid points can be eliminated from consideration because they overproduce/underproduce the absorption in the various other transitions at the velocity of that component. For example, for some components the ionization parameter, $U$, can be tuned to match the observed absorption in {\FeII} or other low ionization transitions, while in others it can be tuned to fully produce the observed {\SiIII} absorption.  Similarly, the metallicity, $Z$, can be tuned to match the observed absorption in the higher order Lyman series lines.  If $Z$ is tuned to match the \lya\ profile then the higher order Lyman series lines would be severely overproduced. This places a lower limit on the metallicity for a given component.

For this absorption complex, the low-ionization \MgII\ bearing phase alone cannot account for the detected higher-ionization transitions (e.g., \CIV, \NV, \OVI) or the \lya. Therefore, another phase, presumably with higher ionization parameter, is introduced into the model. The \NV\ profile is the 'cleanest' among the observed high-ionization transitions over most of this absorption complex, i.e. least saturation and no blends. Thus the Voigt profile fit parameters for \NV\ line, i.e. $\log N(\NV)$ and $b(\NV)$, serve as the starting point for our {\sc cloudy} models of the high-ionization phase. This procedure follows the same steps as those for the low ionization phase, constraining the $U$ and $Z$ of the individual components, with both the low ionization and the high ionization phases combined in order to synthesize model profiles for comparison to the observed profiles. Below we explain the modeling method in more detail in the context of our presentation of the constraints on model parameters for systems A, B, and C. The complete set of absorption lines along with the synthetic model profiles arising from the systems A \& B and system C are shown in Fig. 2 and Fig. 3, respectively.      
 
We continue to treat the subsystems separately as A, B, C because the velocity spread is quite large to have been produced by a single galaxy, and the metal lines and high order Lyman lines separate into these groups, but note that this is not entirely physically motivated.

%==================================================================================
%\setcounter{figure}{1}
\begin{subfigures}
\begin{figure*} 
\includegraphics[width=0.85\textwidth,angle=00]{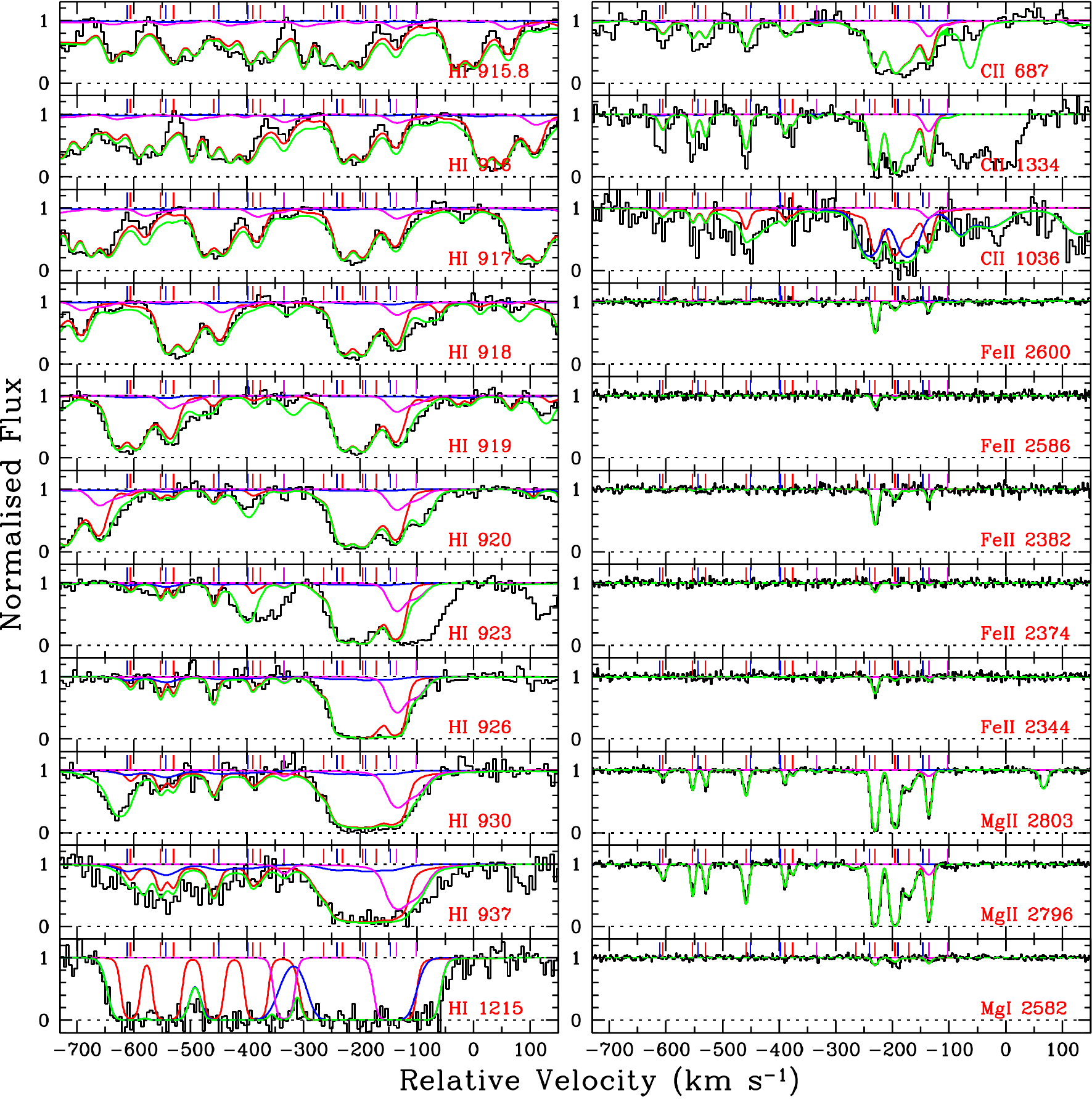} 
\vskip-0.1cm  
\caption{Velocity plots for the systems A and B. The zero velocity corresponds to the galaxy redshift 
of $z_{\rm gal} = 0.9289$. System-A spans from $\sim-700$ to $-300$ \kms\ and the rest is system-B. 
The synthetic profiles corresponds to our adopted photoionization models for the low- (red), 
intermediate- (magenta), and high- (blue) ionization phases summarized in Table~\ref{tab:PImodel}. The 
resultant model profiles are shown in green. The positions of the low-, intermediate-, and high-ionization 
absorption line components are indicated by the vertical tick marks with corresponding colors.}    
\label{fig_modelABa}  
\end{figure*} 
%================================================================================== 
%\setcounter{figure}{1}
\begin{figure*} 
\includegraphics[width=0.85\textwidth,angle=00]{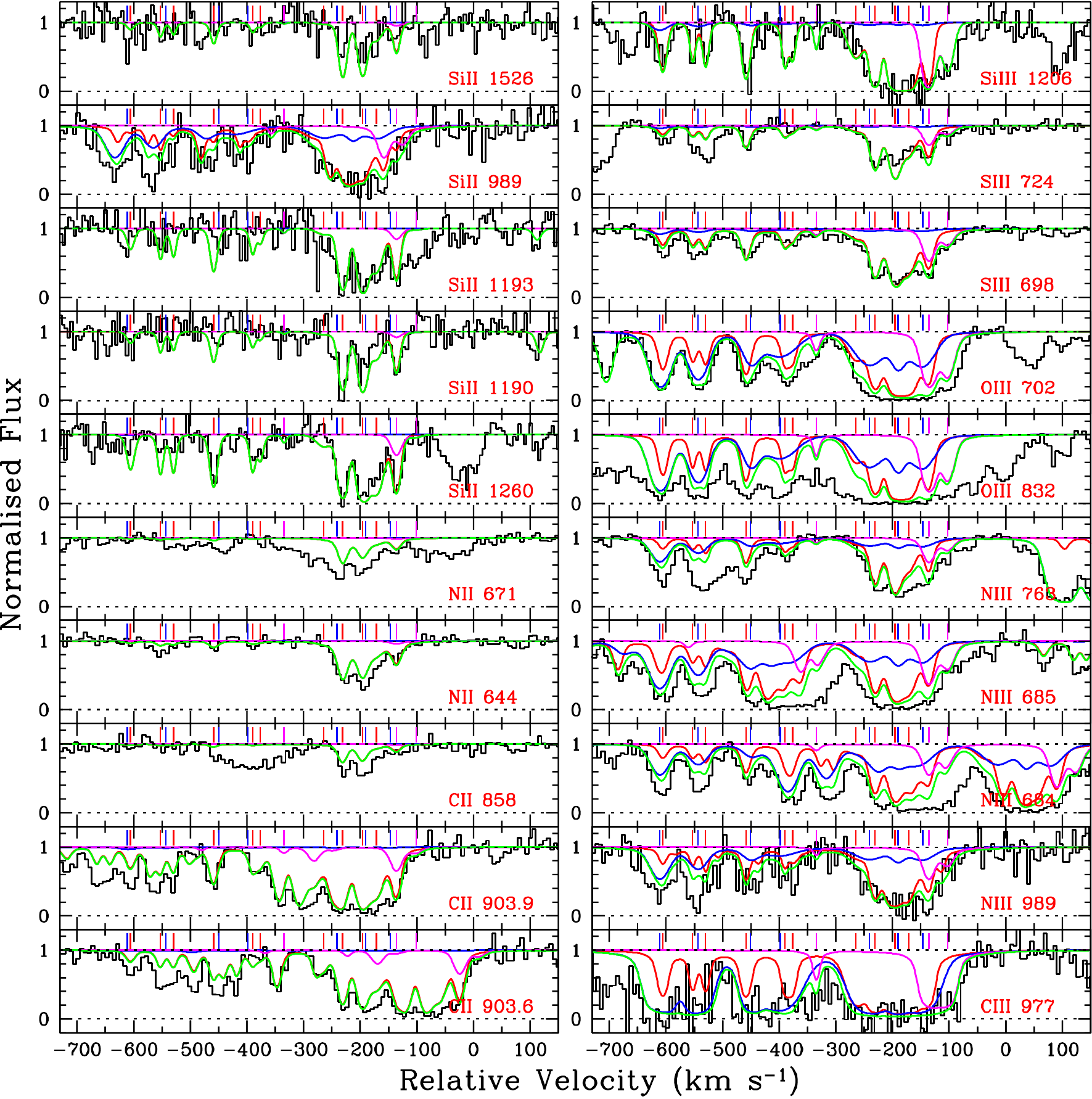} 
\caption{Continued.}    
\label{fig_modelABb}  
\end{figure*} 
%================================================================================== 
%\setcounter{figure}{1}
\begin{figure*} 
\includegraphics[width=0.85\textwidth,angle=00]{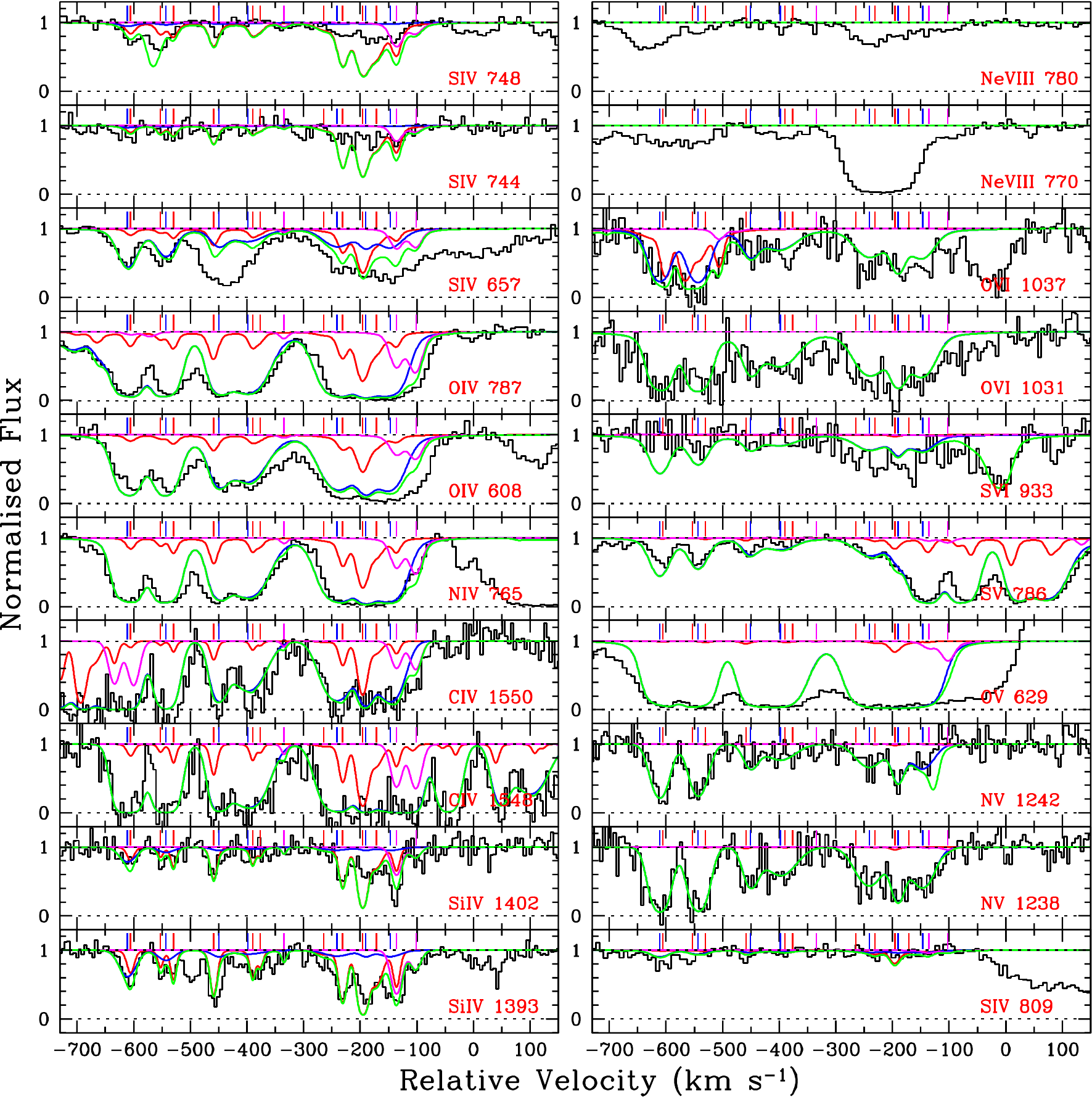} 
\caption{Continued.}    
\label{fig_modelABc}  
\end{figure*} 
\end{subfigures}
%==================================================================================
%================================================================================== 
%\setcounter{figure}{2}
\begin{subfigures}
\begin{figure*} 
\centerline{\vbox{ 
\centerline{\hbox{ 
\includegraphics[width=0.85\textwidth,angle=00]{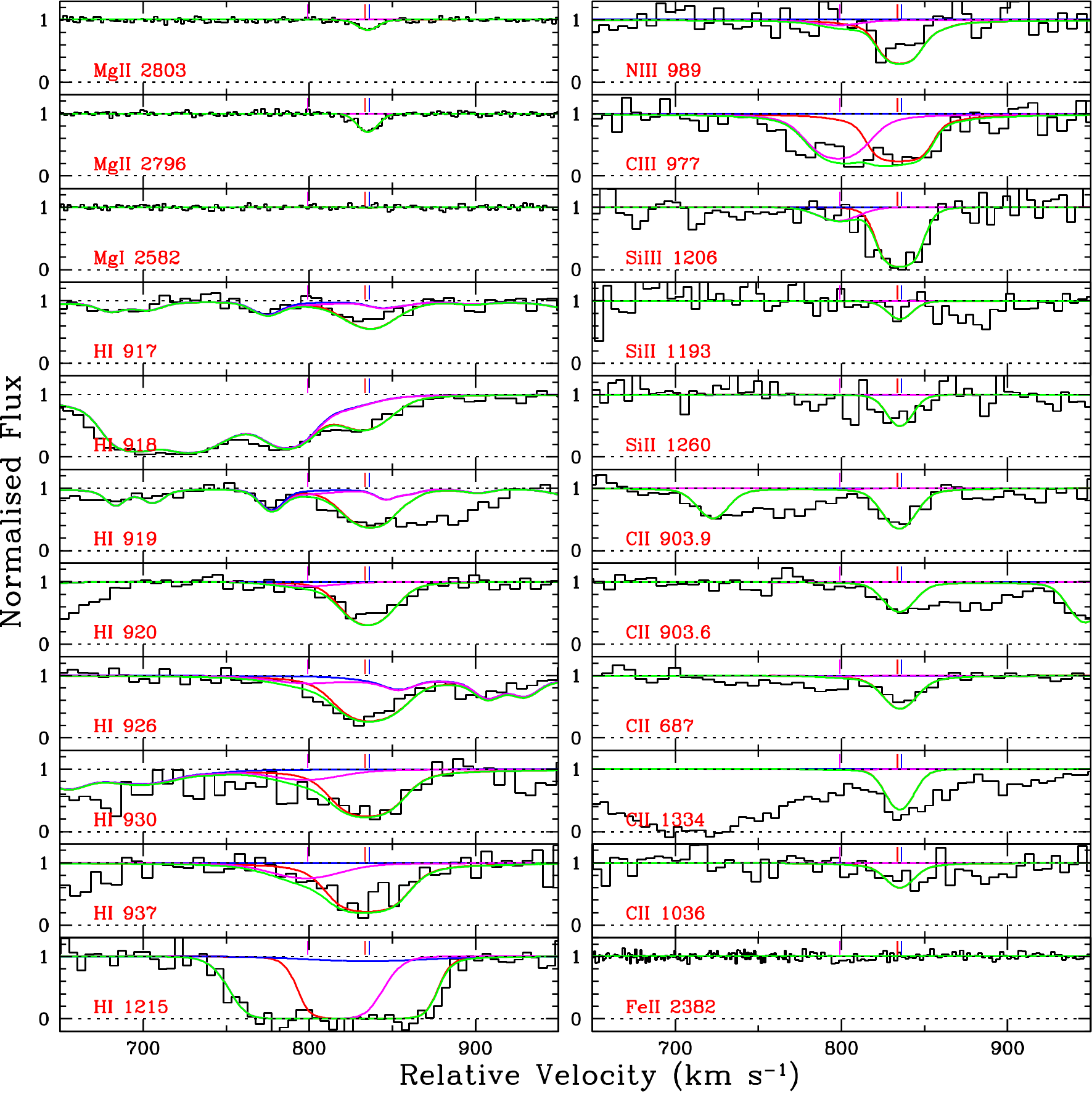}  
}} 
}}  
\caption{Same as Fig.~\ref{fig_modelABa} but for System C.}    
\label{fig_modelCa}  
\end{figure*} 
%==================================================================================
%================================================================================== 
%\setcounter{figure}{2}
\begin{figure*} 
\centerline{\vbox{ 
\centerline{\hbox{ 
\includegraphics[width=0.85\textwidth,angle=00]{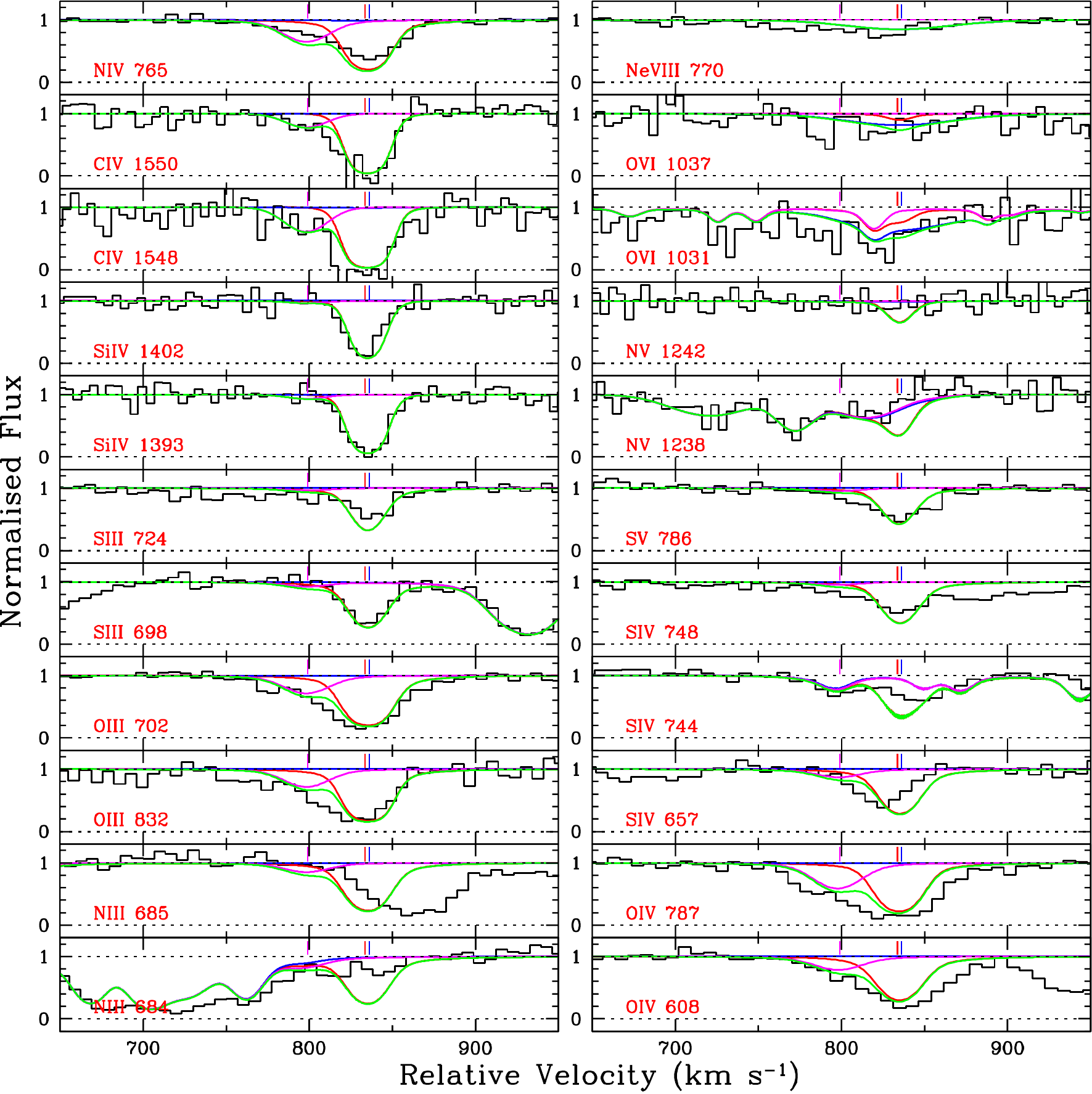}  
}} 
}}  
\caption{Continued.} 
\label{fig_modelCb}  
\end{figure*} 
\end{subfigures}
%==================================================================================  

%==============================================================================================
\begin{table*}
\begin{center}
\caption{PI model parameters for systems A, B, and C}  
\begin{tabular}{ccccccccccccc}  
\hline 
System  & Cloud ID & $v$ &  $\log N(\HI)$ & $b(\HI)$  & $\log N(X)$  &  $b(X)$  &  $\log Z$  & $\log U$  & $\log n_{\rm H}$  & $\log N_{\rm H}$ &  
Thickness & $T$ \\ 
        &       & (\kms)  & (cm$^{-2}$) & (\kms)  & (cm$^{-2}$) & (\kms) & ($Z_{\odot}$) &  & (cm$^{-3}$) & (cm$^{-2}$) & ($\rm kpc$) & ($\rm K$) \\    
\hline      
A &\MgII{} 1  & $-$605   & 14.7  &   8.7  &  11.9 &6.0    & $+$0.5  & $-$2.4  &   $-$2.6  &  17.1    & 0.014 &   2490  \\   
  &\MgII{} 2  & $-$552   & 14.9  &   6.3  &  12.2 &2.9    & $+$0.5  & $-$2.6  &   $-$2.4  &  17.1    & 0.009 &   1960  \\   
  &\MgII{} 3  & $-$529   & 14.9  &   7.0  &  12.1 &3.0    & $+$0.5  & $-$2.4  &   $-$2.6  &  17.3    & 0.022 &   2490  \\   
  &\MgII{} 4  & $-$458   & 15.2  &   7.9  &  12.4 &4.8    & $+$0.5  & $-$2.4  &   $-$2.6  &  17.6    & 0.047 &   2470  \\
  &\MgII{} 5  & $-$390   & 14.8  &   7.0  &  12.0 &3.1    & $+$0.5  & $-$2.4  &   $-$2.6  &  17.2    & 0.017 &   2490  \\
  &\MgII{} 6  & $-$376   & 14.4  &   7.3  &  11.6 &3.7    & $+$0.5  & $-$2.4  &   $-$2.6  &  16.8    & 0.007 &   2500  \\ \\
  &\NV{} 1    & $-$610   & 14.5  &  21.5  &  14.3 &18.0   & $+$0.5  & $-$1.2  &   $-$3.8  &  18.5    &   5.8 &   9070  \\  
  &\NV{} 2    & $-$542   & 14.8  &  24.3  &  14.3 &18.1   &    0.0  & $-$1.2  &   $-$3.8  &  18.9    &  16.3 &   17100 \\
  &\NV{} 3    & $-$450   & 14.4  &  22.2  &  13.7 &15.7   & $-$0.2  & $-$1.3  &   $-$3.7  &  18.4    &   3.8 &   16200 \\
  &\NV{} 4    & $-$398   & 14.4  &  37.9  &  13.8 &34.5   &    0.0  & $-$1.3  &   $-$3.7  &  18.5    &   4.6 &   16200 \\ \\
  &\lya{} 1   & $-$334   & 14.5  &  10.0  &  ---  &10.0   &    0.0  & $-$2.5  &   $-$2.5  &  17.2    & 0.014 &   9220  \\
\hline                                                                                         
B &\MgII{} 7  & $-$264   & 15.2  &  20.5  &  11.7 &16.2   & $-$0.2  & $-$2.9  &   $-$2.1  &  17.5    & 0.012  &   9890  \\  
  &\MgII{} 8  & $-$231   & 16.6  &  12.2  &  13.4 &5.7    &    0.0  & $-$3.0  &   $-$2.0  &  18.8    & 0.170  &   7290  \\  
  &\MgII{} 9  & $-$196   & 16.5  &  13.5  &  13.3 &7.4    &    0.0  & $-$2.7  &   $-$2.3  &  19.0    & 0.583  &   8070  \\  
  &\MgII{} 10 & $-$171   & 15.8  &  17.0  &  12.6 &12.7   &    0.0  & $-$2.8  &   $-$2.2  &  18.2    & 0.077  &   8050  \\
  &\MgII{} 11 & $-$136   & 16.2  &  13.2  &  12.8 &5.1    & $-$0.2  & $-$3.0  &   $-$2.0  &  18.5    & 0.090  &   9420  \\   \\
  &\NV{} 5    & $-$241   & 14.6  &  33.8  &  14.0 &29.9   &    0.0  & $-$1.3  &   $-$3.7  &  18.6    &  6.3   &   16200  \\  
  &\NV{} 6    & $-$190   & 14.4  &  20.5  &  13.9 &12.4   &    0.0  & $-$1.2  &   $-$3.8  &  18.5    &  6.7   &   17300  \\
  &\NV{} 7    & $-$147   & 14.5  &  29.7  &  13.9 &25.1   &    0.0  & $-$1.3  &   $-$3.7  &  18.6    &  5.9   &   16200  \\   \\ 
  &\SiIV{} 1  & $-$136   & 15.6  &  17.3  &  12.6 &10     & $-$0.3  & $-$2.5  &   $-$2.5  &  18.4    &  0.238 &   12500   \\
  &\SiIV{} 2  & $-$102   & 15.3  &  20.5  &  13.1 &10     & $-$0.8  & $-$2.1  &   $-$2.9  &  18.6    &  1.0   &   20100  \\
\hline                                                                                        
C &\MgII{} 12 & $+$835   & 16.1  &  16.4  &  12.1 &7.5    & $-$0.2  & $-$2.1  &   $-$2.9  &  19.3    & 5.5    &   13400  \\ \\         
  &\CIV{} 1   & $+$798   & 14.9  &  25.2  &  13.3 &15.3   & $-$1.0  & $-$1.9  &   $-$3.1  &  18.6    & 1.45   &   23500  \\ \\ 
  &\OVI{} 1   & $+$835   & 12.7  &  49.1  &  14.0 &40.0   &    0.0  &    0.0  &   $-$5.0  &  18.5    & 100.2  &   52700  \\    
\hline
\label{tab:PImodel}          
\end{tabular}
\end{center} 
\end{table*}

%============================================================================================

\subsection{Results for System A}  

\subsubsection{{Low-ionization Phase (\rm ``\MgII-phase'')}}   

The Voigt profile fits of the six \MgII\ clouds (components) in system A, used as constraints for our photoionization model, are from \cite{Churchill1999} and are listed in Table~\ref{tab:PImodel}, and plotted in the bottom right of Figure~\ref{fig_modelABa}. The metallicities of the clouds are individually constrained as $\log Z \sim +0.5$, in order to match the profiles of the higher order Lyman series lines (left panel).  The uncertainty in this determination, within the context of the model assumptions (e.g. uniform density, slab geometry, solar abundance pattern, etc.) is $\sim$0.3~dex. A substantially lower metallicity for any of the clouds would substantially overproduce the corresponding \HI\ absorption. This constraint is consistent with the value found by D03 using a low resolution FOS spectrum of the Lyman series. A similar metallicity for the low ionization clouds was also found by T11, also using the medium resolution coverage of the high order Lyman series lines from the COS G160M observations.

\FeII\ is not detected in any of the system A clouds (middle right of Fig.~\ref{fig_modelABa}), so only a lower limit of $\log U \gtrsim -3.1$ could be obtained from the limiting value of $N(\FeII)$, assuming a solar abundance pattern. The upper limits on $\log U$ rely on \SiIV\ as the main constraint (lower left of Fig.~\ref{fig_modelABc}), but the actual values of $\log U$ depend on whether or not the \SiIV\ absorption arises in the low-ionization phase. One cannot rule out the possibility that \SiIV\ stems from the high-ionization gas phase giving rise to \NV\ and/or \OVI, so we investigate both scenarios.

First, assuming that the \SiIV\ is fully produced in the ``\MgII-phase'', we find $\log U$ values of $-2.6 \le \log U \le -2.4$, which represent upper limits on $\log U$. These values match the \SiII\ absorption but the unblended \CII~\lam1335 and \lam687 lines from clouds 1--3 are slightly under-produced, suggesting a deviation from a solar abundance pattern (D03, T11). \SiIII~\lam1207 and \SIII~\lam698 are fully produced with these values. Some \SIV, \NIII, \CIII, and \OIII\ are produced in these clouds and very little of the \CIV, \OIV, and \NIV\ indicating this is a multiphase absorption system (see red line in left panel of Fig.\ref{fig_modelABc}). The line of sight thicknesses of the low-ionization clouds range from $\sim$~7--50 $\rm pc$.

The model profile of the low ionization clouds 1 and 4 for \SiIV{}\lam1394 for the above upper limits on $\log U$ are slightly narrower than the observed \SiIV~\lam1394 profile (red line in Fig.\ref{fig_modelABc}). We therefore investigate an alternate scenario where the ``\MgII-phase'' ionization parameters are lower for all six clouds, and the \SiIV{} is produced entirely in the high-ionization phase. The $\log U$ values for the \MgII{} clouds in this case are $-2.9 < \log U< -2.5$. The \SiII{} and \CII{} profiles remain for the most part unchanged, however the \SiIII~\lam1207 and \SIII~\lam 698 are slightly under-produced. There is slightly less \NIII, \CIII, and \OIII, and negligible amounts of \SIV, \CIV, and higher ionization transitions. The higher densities of these clouds lead to smaller sizes than the model with higher ionization parameters, with thicknesses ranging between $\sim$~1--13~$\rm pc$.

Our modeling procedures thus find upper and lower limits for the ionization parameters of the \MgII\ clouds based on whether or not all of the \SiIV\ absorption arises in the same phase. Based on the high ionization phase model, discussed in the following section, we prefer the model in which \SiIV\ is produced in the low-ionization phase with the \MgII. Parameters for this preferred model for clouds 1--6 are given in Table~\ref{tab:PImodel}.

\subsubsection{High-ionization Phase (``\rm \NV-phase'')}   

We performed Voigt profile fits to the \NV~$\lambda\lambda$1238,1242 doublet profiles using {\sc vpfit}\footnote{http://www.ast.cam.ac.uk/$\sim$rfc/vpfit.html} and obtained four \NV\ components (see Fig.~\ref{fig:vpfit} and lower right of Fig.~\ref{fig_modelABc}), like those found by T11. We then optimize on these \NV\ column densities, that is for each of a grid of values of $U$ and $Z$, we find the total hydrogen column density for which the measured \NV\ column density will arise.  We then compare synthesized model spectra to the observed profiles to constrain {\sc cloudy} models of the high ionization phase. The contributions from the low-ionization clouds are combined with those from these ``\NV-phase'' when model profiles are synthesized and compared to the data (blue:high ions only, green:low+high in Fig.~\ref{fig_modelABa},~\ref{fig_modelABb},~\ref{fig_modelABc}).

With the ionization parameters set to give the maximum \OVI\ absorption that could be consistent with the data, the model comes close to accounting for the saturated \CIV\ absorption (Fig.~\ref{fig_modelABc}). Both \CIVdblt\ doublet members, saturated in the data, are reproduced in \NV\ clouds 1 and 2 with these values. The \CIV\ in clouds 3 and 4 are slightly under-produced, which can be alleviated by lowering the ionization parameter by a few tenths of a dex, a model that is still consistent with the \OVI\ profiles. The \OIII~\lam\lam702, 832 profiles are slightly under-produced by these higher ionization cloud models, as are the \CIII~\lam977 and \NIII~\lam\lam685,989 profiles (Fig.~\ref{fig_modelABb}).

We prefer models that have values of the ionization parameters slightly lower than the maximum values permitted so as not to exceed the observed \OVI\ absorption, and to better match all the observed transitions. The best match to the data has $\log U \sim -1.2$ for \NV\ clouds 1 and 2 and $\log U \sim-1.3$ for \NV\ clouds 3 and 4. These values account for the strong \CIV\ and other intermediate ionization transitions such as \NIII, \CIII, and \OIII, and simultaneously produce absorption in the higher ionization transitions such as \OVI.  \SV~\lam786 and \SVI~\lam933 are slightly overproduced with these values, however these ionization parameters are not high enough for \NeVIII\ to be produced, and thus any absorption in \NeVIII\ at this velocity {\em must} trace yet another higher ionization gas phase.

The metallicity of \NV\ cloud 1 is effectively constrained by the blue \lya\ wing (bottom left of Fig.~\ref{fig_modelABa}), which is not fit by the ``\MgII-phase'' (red line) and thus we propose arises in the same phase as the \NV. A super-solar metallicity of $\log Z = +0.5$ is needed to fit this wing with the $\log U$ set to $-1.2$. Since they are not well constrained, we initially assume that the metallicities of each the \NV\ clouds in system A are similar, and set the other three cloud's metallicities to $\log Z = 0.5$, as well. However, with these values the \lya\ absorption at $\sim-500$~\kms\ is under-produced, as well as the \HI~\lam937. In order to match these two absorption profiles, the metallicities of \NV\ clouds 2 and 3 were both lowered to $\log Z = 0.0$ and $-0.2$, respectively. With $\log Z = +0.5$ for \NV\ cloud 4, there is substantial unaccounted \lya\ absorption from $-350 \lesssim v \lesssim -300$~\kms. By lowering the metallicity of the \NV\ cloud 4 to $\log Z\sim-1.0$, the unaccounted for \lya\ absorption can be produced. However, overproduction of the Lyman series in this component occurs at $\log Z<0$, so $\log Z \sim -1.0$ is ruled out, and we set $\log Z = 0$ as our preferred model value. The unaccounted for \lya\ absorption would then be traced by another component which we now introduce. The line of sight thicknesses of the high-ionization clouds are on the order of 1--10~$\rm kpc$.

\subsubsection{``{\rm \lya}-only'' Phase}
A low-metallicity, hydrogen cloud is proposed to produce the unaccounted for \lya\ absorption at $\sim-330$~\kms\ (the positions of \NV\ cloud 3 and 4), as discussed above. A cloud with column density of $\log N(\HI) = 14.5$ and a Doppler parameter of $b = 10$~\kms\ provides an adequate fit. Such an additional cloud would not be surprising since its properties are similar to the \MgII\ clouds (see Table~\ref{tab:PImodel}). The metallicity must be around solar to avoid higher order Lyman series overproduction. An ionization parameter of $\log U = -2.0$ overproduces the high-ionization transitions such as \CIV. A lower value of $\log U \sim -2.5$ does not overproduce any transitions. The thickness of this cloud is $\sim$14~$\rm pc$. We plotted this component as a pink line in Fig.~\ref{fig_modelABa} and in Fig.~\ref{fig:modelsum}, the same color as the ''intermediate phase'' introduced in the next section.

\subsection{Results for System B}

\subsubsection{Low-ionization Phase (``\rm \MgII-phase'')}      

Again we begin with the column densities and Doppler parameters for the five \MgII\ components found by \citet[][\MgII\ Clouds IDs 7--11 in Table~\ref{tab:PImodel}; bottom right of Fig.~\ref{fig_modelABa}]{Churchill1999}. The three strongest components, at $v = -231$, $-196$, and $-136$~\kms, were constrained by \FeII\ detections (Fig.~\ref{fig_modelABa}) to have $\log U = -3.0$, $-2.7$, and $-3.0$, respectively, assuming a solar abundance pattern. They provide reasonable match to the \SiII, and \CII, \SiIII, and \SIII\ absorption(Fig.~\ref{fig_modelABb}). However the \SiIV, particularly \SiIV~$\lambda$1402 in the $v = -196$~\kms\ cloud, is overproduced (Fig.~\ref{fig_modelABc}). Visual inspection of the \SiIV\ doublet of this component reveals that the shapes of the two doublet members do not match, a confusion possibly caused by noise or an unidentified blend. However, this would not resolve the discrepancy, since the $\log U = -2.7$ model overproduces {\SiIV}.  The \SiIV\ in the $v = -136$~\kms\ cloud is under-produced (see red line on Fig.~\ref{fig_modelABc}), the opposite situation from the $v = -196$~\kms\ cloud. D03 proposed a solution that introduced an additional intermediate-ionization phase superposed on the \MgII\ cloud to account for this \SiIV\ absorption. The constraints on such a cloud will be discussed further below.  The two other (weaker) clouds, at $v = -171$~\kms\ and $-264$~\kms, do not have \FeII\ detected and therefore were constrained by other low- and mid-range ionization transitions, consistent with the lower limits on $\log U$ placed by the absence of \FeII. The cloud at $v = -264$~\kms\ was constrained to have $\log U \leq -2.9$, since this fits the \CII\ transitions and does not overproduce \SiIII~$\lambda$1207. The cloud at $v = -171$~\kms\ was constrained to have $\log U \sim -2.8$, as higher values overproduce \SIII~$\lambda$698 and \SiIV~$\lambda$1394 and lower values do not produce enough \SIII~$\lambda$698. These values are similar to those derived for the stronger \MgII\ clouds with \FeII\ detections.

The metallicities for the system B \MgII\ absorbing clouds were constrained by the higher order Lyman series lines covered by COS (left panel, Fig.~\ref{fig_modelABa}). The metallicity values obtained here are in the range $-0.2 \leq \log Z \leq0$, consistent with previous values from D03, who had only FOS coverage of these lines. T11, using COS data, constrained the cloud at $v = -231$~\kms to have $\log Z = -0.3$, which is consistent with our value of $-0.2$, within uncertainties and differences in the EBR.  The thicknesses of the clouds range from 12~$\rm pc$ for the smallest cloud to 500~$\rm pc$ for the largest cloud.  The properties of the model clouds are summarized in Table~\ref{tab:PImodel}.  We note that clouds \MgII~8 and 9 have the largest $N{(\HI)}$ and they dominate in giving rise to the observed partial Lyman limit break (D03).

The \CIII\ and \OIII\ lines saturate with the adopted $\log U$ and $\log Z$ values in the clouds near $v \sim -250$~\kms, however, the absorption in the wings of these lines is not fully produced by the \MgII\ clouds (red lines in Fig.~\ref{fig_modelABc}). Furthermore, \CIV\ and higher ionization species are not fully produced by the relatively low-ionization \MgII\ clouds. Evidently, system B requires a higher ionization phase, as was also needed to explain all the absorption lines in system A.

\subsubsection{High-ionization Phase (``\rm \NV-Phase'')}

\NV\ absorption from system B was decomposed with three Voigt profile components at $-241$, $-190$, and $-147$~\kms\ (see Table~\ref{tab:PImodel}, Fig.~\ref{fig_modelABc}). We first seek to find a model which produces the \CIV, other intermediate ionization transitions, and the \OVI. Ionization parameters of $\log U = -1.3$, $-1.2$, $-1.3$ for the three \NV\ clouds reproduce the saturated \CIV\ as well as the strong \OVI. The other intermediate transitions are explained well with these parameters, with the exception of \SVI\, which is under-produced. This contrasts with the overproduction of \SVI\ in the system A high ionization phase. \NeVIII\ is not produced for these parameters, and remains unproduced even when the ionization parameters are raised to $-1.0$, which exceeds the observed \OVI. Therefore, we prefer a model where the detected \NeVIII\ (see T11) traces a higher ionization phase, similar to our preference for the system A model.

The metallicities of the \NV\ clouds are not well constrained by the data. The best constraint available is to avoid overproduction of the blue side of the Lyman series lines at $v \sim -250$~\kms\ for cloud \NV~5. A metallicity of $\log Z \sim -0.5$ slightly overproduces the \HI~\lam\lam938, and 931 profiles, while a solar abundance provides an adequate fit. The metallicities of \NV\ clouds 6 and 7 are not constrained by the data since they fall in the middle of the \lya\ profile and do not produce Lyman series absorption. We adopt the same metallicity, $\log Z \sim 0.0$, for all three \NV\ clouds under the assumption that the \NV\ clouds trace similarly enriched gas, with the caveat that these are rather arbitrary and uncertain values.  At this metallicity, the line of sight thicknesses of these three \NV\ clouds fall between $\sim$5.9 and 6.7~$\rm kpc$.

\subsubsection{Intermediate Phase (``\rm \SiIV-Phase")} 

The \MgII\ cloud 11, constrained by \FeII, does not entirely account for the \SiIV\ absorption at $v = -136$~\kms (red line, Fig.~\ref{fig_modelABc}). \NIII~$ \lambda$989 is also somewhat underproduced by cloud \MgII\ $11$ alone at this velocity (lower right of Fig.~\ref{fig_modelABb}). D03 introduced a mid-ionization cloud superposed on the \MgII\ cloud at this velocity. Adopting the column density and $b$-parameter from the cloud in their paper, values of $\log U = -2.5$ and $\log Z = -0.3$ provide a good match to the data. This cloud's thickness is $\sim$240~$\rm pc$.

In a number of transitions, notably \lya, \HI~$\lambda$930, and \HI~$\lambda$926 there is absorption at $-102$~\kms\ that is not produced by any of the above clouds. This absorption is also seen in the \SiIII~$\lambda$1207, \SIII~$\lambda$698, \CIII~$\lambda$977, \CIV~$\lambda$1551, \OIII~$\lambda$702, \OIV~$\lambda$608 and $\lambda$787, and the \NIV~$\lambda$765 profiles. There also is a weak component in the \SiIV~\lam1394 profile at this velocity, which is consistent with the \SiIV~\lam1403 profile, within the noise.  A cloud was therefore added to the {\sc cloudy} model optimized on a weak \SiIV\ component with $\log N(\SiIV)=13.1$.

The metallicity of the added cloud should be low enough to produce the red wing of the \lya\ as well as the \HI~$\lambda$931 and $\lambda$926 lines, yielding $\log Z = -0.8$. With this metallicity the ionization parameter is constrained to produce as much of the missing absorption as possible. A value of $\log U = -2.1$ produces a good match to the \SIII~$\lambda$698, \CIII~$\lambda$977, \CIV~$\lambda$1551, and \OIV~$\lambda$787. \SiIII~$\lambda$1207, \OIII~$\lambda$702, and \OIV~$\lambda$608 are not overproduced, and \NIV~$\lambda$765 is slightly overproduced. Despite these minor discrepancies, it seems clear that a cloud with similar properties to this is needed. The thickness of the cloud turns out to be $\sim$1.6~$\rm kpc$.

\subsection{Results for System C}

\subsubsection{Low-ionization Phase (``\rm \MgII-Phase")}    

System C is a single-cloud, weak \MgII\ absorbing system at $v=+835$~\kms (top left, Fig.~\ref{fig_modelCa}). There is also an offset cloud at $v=+798$~\kms\ that is apparent in the \CIV\ profile (top left, Fig.~\ref{fig_modelCb}), and which does not give rise to low ionization absorption. For the \MgII\ cloud, D03 considered both a one-phase model, where the \OVI\ is produced together with the \MgII, and a two-phase model. We begin by testing their one-phase model in light of the higher resolution data now available.

For $\log U = -1.9$, \OVI\ is produced with $\log N(\OVI)\sim$14. With this ionization parameter, a metallicity of $\log Z = 0.1$ is necessary to fit the high order Lyman series, however the \lya\ is under-produced. In order to match the \lya\ profile, a lower metallicity of $\log Z = -0.1$ is needed, however for this metallicity the Lyman series is overproduced. Furthermore, while the $\rm Si$ and $\rm O$ ions are fit well, the \SV~\lam786 line is overproduced, as are the nitrogen lines \NIII~\lam\lam685, 989, \NIV~$\lambda$764, and \NV~$\lambda\lambda$1238,1242. The \NV~$\lambda$1238 line is blended with \NV~$\lambda$1243 from system B, but \NIII\ and \NIV\ can be used to constrain this possible nitrogen deficiency. Nitrogen would be deficient by about one dex if this single-phase, $\log U = -1.9$ model is to be correct. The inability to simultaneously reproduce the \lya\ and Lyman series lines, and the overproduction of \SV~\lam786 and nitrogen, suggest that we consider a lower ionization model.

For this lower ionization model, the ionization parameter still has to be at least $\log U \geq -2.1$ to account for the \SiIV (middle left, Fig.~\ref{fig_modelCb}). At this ionization parameter a metallicity of $\log Z = -0.2$ best explains to the \lya\ and other Lyman series profiles (left, Fig.~\ref{fig_modelCa}). Some of the higher order series lines are slightly overproduced, however lowering the metallicity would result in \lya\ under-production. With these parameters, the \SV~$\lambda$786 (middle right, Fig.~\ref{fig_modelCb}) is no longer overproduced and all other ions have adequate profile match. The nitrogen ions model profiles are still slightly stronger than the data, but only a $\sim$0.1 dex decrease in the abundance of N would be needed for consistency. This model cloud produces an \OVI\ column density of $\log N(\OVI) = 13.2$, and thus the remaining \OVI\ column must reside in an additional, separate slightly lower density phase of gas. The line of sight thickness of this cloud is 5.5~$\rm kpc$.

\subsubsection{Intermediate Phase (``\rm \CIV-phase'')}   

A second component, not observed in \MgII\ or in the higher order Lyman series, is necessary to account for the blue-ward \lya\ absorption at 750 $<$ v $<$ 800~\kms\ not accounted for by the \MgII\ component (bottom left, Fig.~\ref{fig_modelCa}). This component is also seen in several metal-line profiles starting with \CIII~$\lambda$977 (top right, Fig.~\ref{fig_modelCa}) and is clearly seen in the \NIV~$\lambda$764  profile and  \CIVdblt\ profiles(top left, Fig.~\ref{fig_modelCb}). Due to noise, it is not clear whether the \OVI\ absorption (top right, Fig.~\ref{fig_modelCb}) also exhibits this asymmetry in its profile, although visual inspection indicates the possibility. \NV~\lam1243 is not detected. Adopting the Doppler parameter and column density from D03 of the offset \CIV\ absorption, we constrain the ionization parameter and metallicity of this offset cloud. The ionization parameter is constrained mainly by the other intermediate ionization transitions, such as \CIII~\lam977(Fig.~\ref{fig_modelCa}), \OIII~\lam\lam702, 832, \SV~\lam786, and \OIV~\lam608, \lam787 (Fig.~\ref{fig_modelCb}). The best match model is produced with an ionization parameter of $\log U = -1.9$. The $\log Z$, constrained by the \lya, is found to be $-1.0$. Lower values overproduce the \HI~\lam938 and higher order Lyman series lines. The thickness of this cloud is $\sim$1.4~$\rm kpc$, similar to the \MgII\ cloud.

\subsubsection{High-ionization Phase (``\rm \OVI-phase'')}   

With the above two clouds, all the absorption is accounted for besides the majority of the \OVI\, and \NeVIII. We therefore add to the model an \OVI\ component with $b = 40$~\kms\ and $\log N(\OVI) = 14.0$. The ionization parameter (density) must be high (low) for \NeVIII\ to be photoionized, which constrains the value of $\log U$ to be $0.0$, corresponding to a density of $\sim 10^{-5}$~cm$^{-3}$. In order that the thickness of this cloud is not unrealistically large ($\sim$1 Mpc, larger than the halo itself), the metallicity must be near to or exceed solar; a value of $\log Z = 0.0$ gives a line of sight thickness of 100~$\rm kpc$ and does not exceed the observed \lya\ absorption. It appears from the data that \NV\ is not detected, and thus the \OVI\ and \NeVIII\ trace the same phase of gas, which if photoionized, is high metallicity gas. The alternative possibility of a hotter, but higher density, collisionally ionized cloud producing the observed \OVI\ and \NeVIII. Such models are explored in Section~\ref{sec:CI}.

\subsection{Effects of Alternative EBR}	  
\label{alterEBR}  

In order to consider the uncertainties in model parameters, we repeated our analysis using, instead of HM01, the recent UV background spectrum published by \citet[][hereafter KS15]{Khaire2015}, normalized at the redshift of this system.  

Our conclusions do not change qualitatively, but the parameters of the simplest suitable model do change. The metallicities of the low ionization clouds needed to be adjusted upward up to by $\sim$0.3--0.5~dex with the KS15 EBR, in order that the Lyman series would not be overproduced, but the similar $\log U$ values are still suitable. For the higher ionization clouds, the ionization parameters for the KS15 EBR fit are $\sim$0.5~dex lower than for the HM05 model. Although these differences are not completely trivial, they would not change our overall conclusions. To put things in perspective, the metallicity and density can have values ranging over several orders of magnitude. An uncertainty of a factor of 2 or 3, due to uncertainties in the EBR, is not very significant relative to the range of possibilities.   

We also note that the inclusion of galaxy G2's stellar radiation field does not substantially alter the results of our photoionization models. We refer the reader to section 6.4 of D03 for a detailed calculation, as well as Appendix B of \cite{Churchill1999}.

%============================================================================================
\begin{table} 
\begin{center} 
\caption{Voigt profile fit parameters for \CIV, \NV, and \OVI.}                  
\begin{tabular}{lcccc}  
\hline 
Ion  &     $v$ (\kms)    &  $b$ (\kms) &  $\log N$ (cm$^{-2}$) \\    
\hline  
 \NV  &  $-613$  & 18.4 $\pm$ 1.7  & 14.35 $\pm$ 0.06  \\   
\CIV  &          &                 & 14.60 $\pm$ 0.19  \\  
\OVI  &          &                 & 14.99 $\pm$ 0.18  \\    
\NeVIII &        &                 & 13.71 $\pm$ 0.29  \\  \\   
 \NV  &  $-540$  & 17.7 $\pm$ 1.6  & 14.33 $\pm$ 0.06  \\   
\CIV  &          &                 & 14.85 $\pm$ 0.20  \\   
\OVI  &          &                 & 14.98 $\pm$ 0.27  \\  
\NeVIII &        &                 & 14.04 $\pm$ 0.08  \\ \\ 
 \NV  &  $-457$  & 14.0 $\pm$ 2.5  & 13.64 $\pm$ 0.08  \\   
\CIV  &          &                 & 14.47 $\pm$ 0.21  \\  
\OVI  &          &                 & 13.77 $\pm$ 0.16  \\    
\NeVIII &        &                 & (not detected)    \\ \\       
 \NV  &   $-392$ & 32.7 $\pm$ 3.4  & 13.81 $\pm$ 0.07  \\
\CIV  &          &                 & 14.47 $\pm$ 0.06  \\      
\OVI  &          &                 & 14.40 $\pm$ 0.06  \\  
\NeVIII &        &                 & 14.07 $\pm$ 0.04  \\  \\  
 \NV  &  $-241$  & 27.2 $\pm$ 3.2  & 13.93 $\pm$ 0.06  \\  
\CIV  &          &                 & 14.58 $\pm$ 0.10  \\      
\OVI  &          &                 & 14.62 $\pm$ 0.07  \\  \\   
 \NV  & $-192$   &  8.7 $\pm$ 3.5  & 13.82 $\pm$ 0.12  \\   
\CIV  &          &                 & 13.69 $\pm$ 0.78  \\      
\OVI  &          &                 & 13.96 $\pm$ 0.38  \\ \\    
 \NV  & $-153$   & 43.2 $\pm$ 6.2  & 14.04 $\pm$ 0.09  \\  
\CIV  &          &                 & 14.80 $\pm$ 0.10  \\      
\OVI  &          &                 & 14.63 $\pm$ 0.05  \\ 
\NeVIII &        &                 & 14.21 $\pm$ 0.05  \\ \\ 
\CIV  & $+830$   & 12.3 $\pm$ 2.8  & 14.73 $\pm$ 0.47  \\   
\OVI  &          &                 & 14.45 $\pm$ 0.15  \\  
\NV   &          &                 & $<$13.5           \\   
\NeVIII &        &                 & 13.90 $\pm$ 0.13  \\     
\hline   
\end{tabular} 
\label{tab:vpfit} 
\end{center}  
~\\ 
Notes-- \NeVIII\ column densities are taken from T11. Note that the COS spectrum could not 
resolve the components at $-241$ and $-192$ \kms. T11 provided an integrated $\log N(\NeVIII)$  
of 14.53$\pm$0.04 dex corresponding to these two components. To be consistent with T11, we have 
added the component column densities of \CIV, \NV\ and \OVI\ for our CIE/non-CIE models. Additionally, 
T11 reported two \NeVIII\ components in system C which are not apparent in any other high 
ionization lines. We, therefore, present the added the individual \NeVIII\ component column densities 
for the $v=+830$~\kms\ component.           
\end{table} 

%============================================================================================
\section{Collisional Ionization Model}
\label{sec:CI}

In this section we explore the viability of equilibrium and non-equilibrium collisional ionization (CIE and non-CIE) models for the high ions, i.e. \CIV, \NV, \OVI, and \NeVIII. Here we adopt the CIE and non-CIE models of \cite{Gnat2007}, in which equilibrium and non-equilibrium cooling efficiencies and ionization states for low density radiatively cooling gas are computed under the assumptions that the gas is optically thin, dust free, and subject to no external radiation field.

\begin{figure} 
\centerline{
\vbox{
\centerline{\hbox{ 
\includegraphics[width=0.50\textwidth,angle=00]{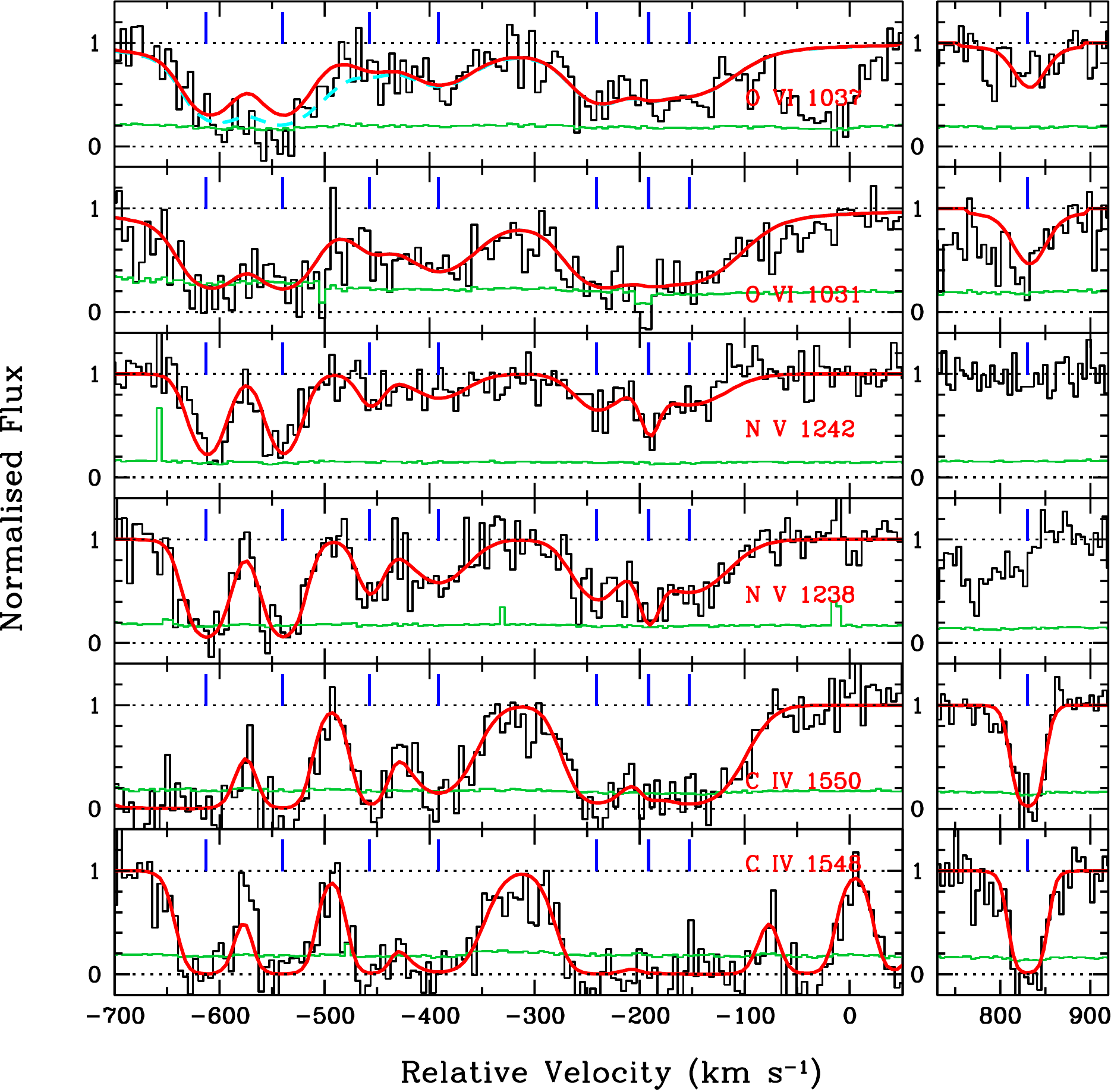}  
}}
}} 

\caption{Voigt profile fits for the high-ionization lines. The zero velocity corresponds to  $z_{\rm gal}$ = 0.9289. The smooth (red) curves are the best-fitting profiles over-plotted  on top of the data (black histogram). \OVI~$\lambda1037$ is blended with \CII~$\lambda1036$  absorption. The cyan (dashed) curve show fit with both \OVI\ and \CII. Errors in each pixels are shown as grey (green) histograms. The ticks represent the line centroids of the  best-fitting Voigt profile components.}    
\label{fig:vpfit}  
\end{figure}

The column densities of high-ions (\CIV, \NV\ and \OVI) in different absorption components were estimated using the {\sc vpfit} software. Both the \CIV\ and \NV\ are covered by the STIS spectrum and hence they are fitted simultaneously assuming pure non-thermal broadening, i.e. $b(\CIV) = b(\NV)$.  For the \OVI\ we have assumed component structure and $b$-parameters similar to \NV. The best fitting Voigt profiles are shown in Fig.~\ref{fig:vpfit} and the fit parameters are summarized in Table~\ref{tab:vpfit}. The \NeVIII\ column densities presented in the table are taken from T11.

%==================================================================================
%==================================================================================

\begin{figure*} 
\includegraphics[width=0.49\textwidth,angle=00]{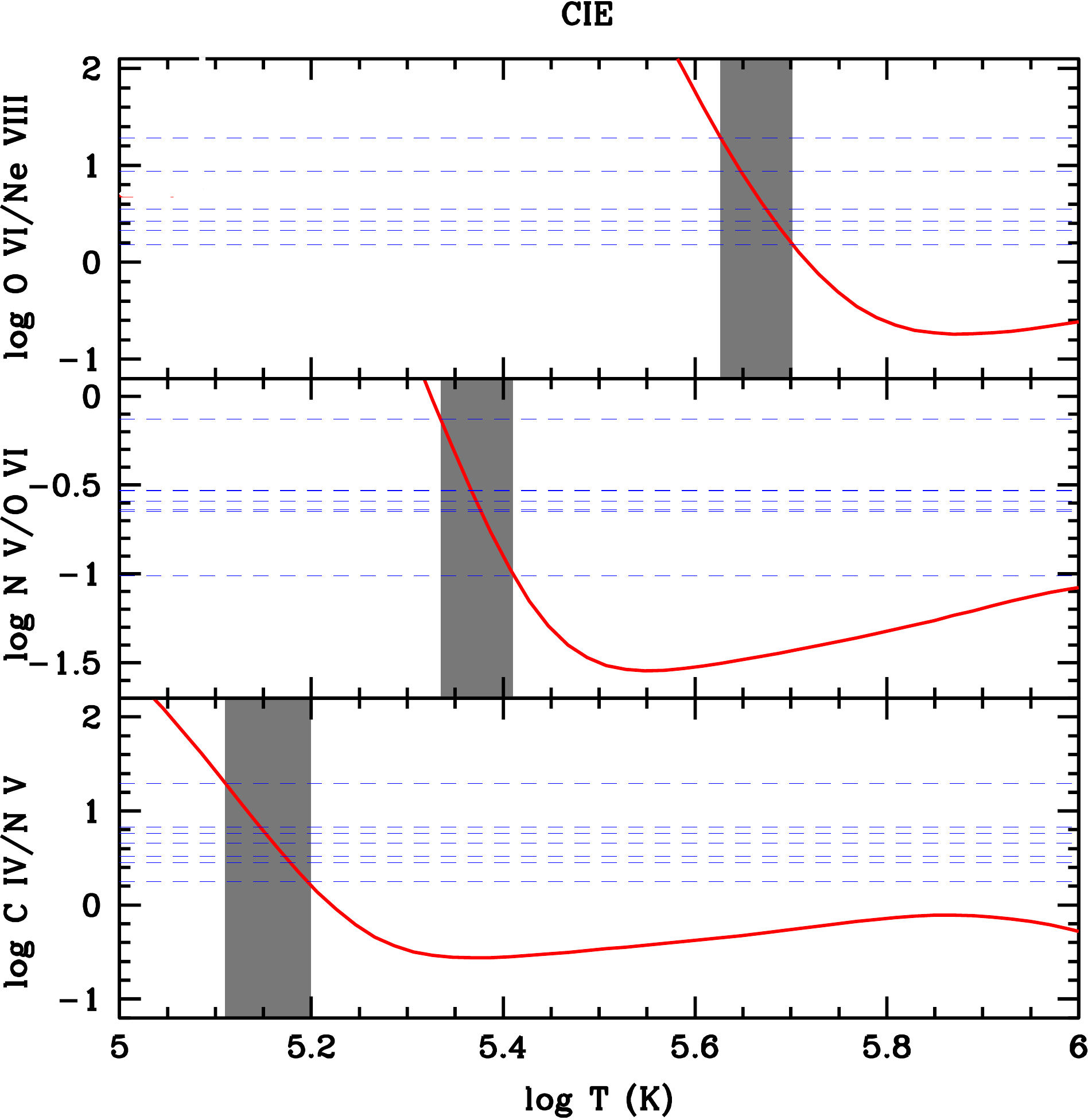}\hfill
\includegraphics[width=0.49\textwidth,angle=00]{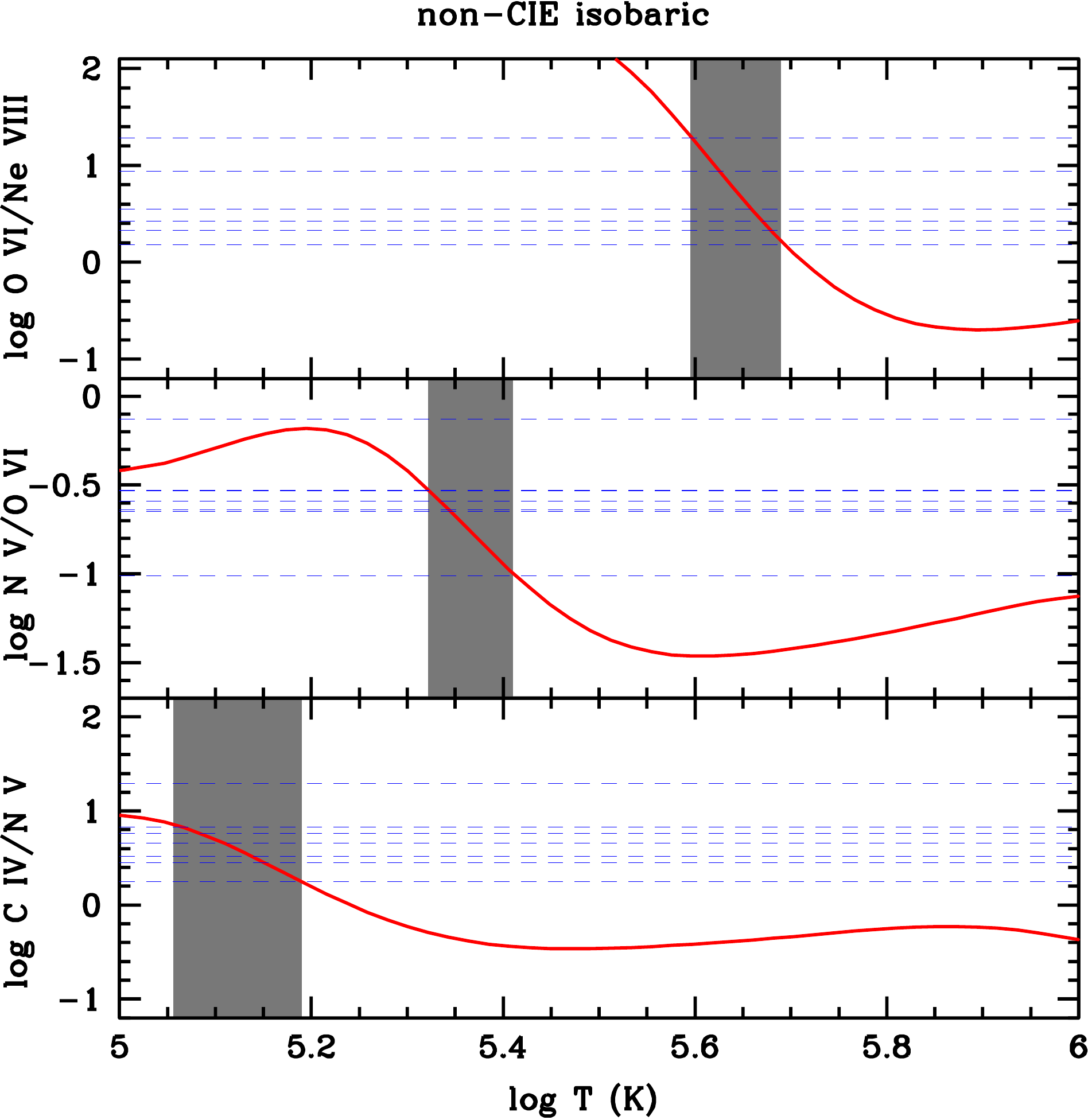}  
\vskip-0.3cm  
\caption{CIE (left) and non-CIE (right) models for the high-ions. Various column density ratios are plotted against gas temperature. The horizontal dashed lines represent the observed values of column density ratios in different absorption components (Table~\ref{tab:vpfit}). The 
red lines are the predicted column density ratios of \citealt{Gnat2007}. The shaded regions indicate overall range in temperature to explain the ratios. No single range in temperature could explain all the ratios simultaneously for any of these models. The straight lines seen in the top left panel is an artifact caused due to inappropriate handling of the \NeVIII\ ionization fraction at temperatures much lower than the temperature at its peak.
}       
\label{fig_CIE}   
\end{figure*}

In Fig.~\ref{fig_CIE} we present CIE and non-CIE models of systems A, B, and C for the high-ions. The different column density ratios (\CIV\ to \NV, \NV\ to \OVI, and \OVI\ to \NeVIII) are plotted against the gas temperature. The horizontal dashed lines represent the observed column density ratios for different components as derived from Table~\ref{tab:vpfit}. For the CIE model, the observed $N(\CIV)/N(\NV)$ ratios are consistent with gas temperature in the range $\log (T/\rm K)\sim$~5.10--5.20. The $N(\NV)/N(\OVI)$ ratios, on the contrary, suggest a very narrow, higher temperature range of $\log (T/\rm K)\sim$~5.35--5.40. Moreover, the $N(\OVI)/N(\NeVIII)$ ratios imply a significantly higher gas temperature, i.e. $\log (T/\rm K)\sim$~5.65--5.70. No single temperature, or range of temperatures, can explain all three ratios simultaneously for any of the components. The non-CIE (both isobaric and isochoric) models also exhibit the same characteristics. Here we only show the isobaric model in the right panel of Fig.~\ref{fig_CIE}. The figure indicates that a single temperature CIE and/or non-CIE model is not suitable to simultaneously reproduce the observed column densities for more than two ions.

In the case of system C, \CIV\ is traced by the \MgII\ cloud and \NV\ is not detected, and thus a CIE/non-CIE isobaric model with $5.6\leq \log T \leq 5.7$ can account for the observed \OVI\ and \NeVIII. If the \OVI\ and \NeVIII\ are photoionized, then a near solar metallicity is required in order not to have an unreasonably large thickness (i.e., $\gtrsim 1$ Mpc, larger than the halo itself). However, since the metallicity is unconstrained we conclude that for system C the collisional ionization model (for \OVI\ and \NeVIII) is an equally feasible scenario.

%==================================================================================
%==================================================================================
% 

\subsection{Comparison with T11 CIE Model}

\begin{figure*} 
\centerline{
\vbox{
\centerline{\hbox{ 
\includegraphics[width=0.50\textwidth,angle=00]{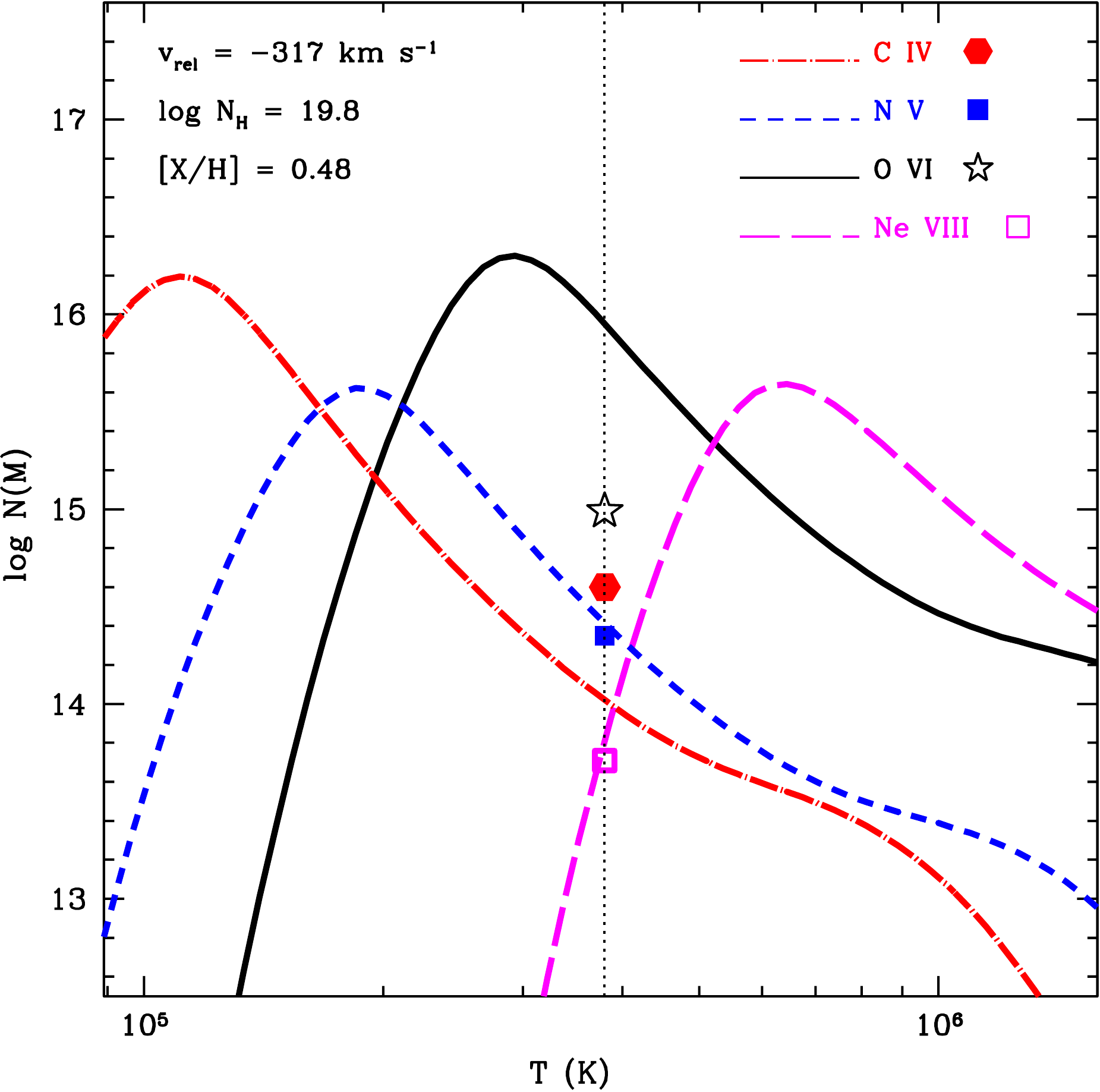}  
\includegraphics[width=0.50\textwidth,angle=00]{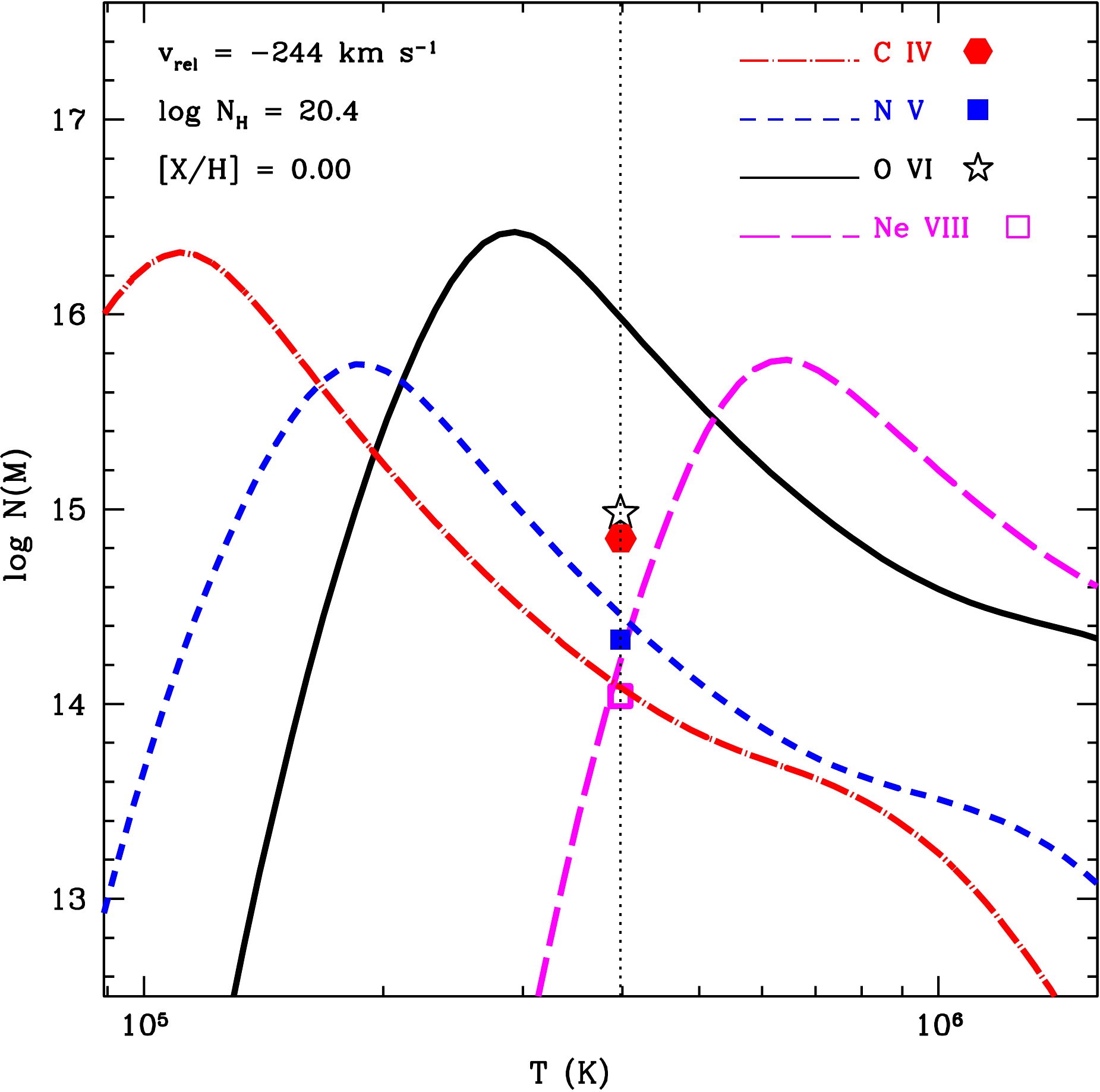}   
}}
}} 
\vskip-0.3cm  
\caption{Revisiting the CIE solutions of T11 for the components at $v = -613$~\kms\ (left) and $-540$~\kms\ (right). Note that these components are at $v=-317$~\kms\ and $-244$~\kms\ in T11 due to a different choice of reference redshift by the authors. In each panel, the model predicted column densities are shown with smooth curves and the observed values are plotted as discrete symbols. Note that the ionic column densities were computed using the $N_{\rm H}$ and $\rm [X/H]$ (or $\log Z$) values as indicated in the plot are obtained by T11. The vertical dotted lines represent their temperature solutions. For both the components the models produce $\log N(\OVI) \sim$~16.0 which is about an order of magnitude higher than the observed values. The model predicted $N(\CIV)$ values, on the other hand, are $\sim0.7$~dex lower than the observed ones.     
}        
\label{fig_compare}   
\end{figure*}

\begin{figure*} 
\includegraphics[width=0.80\textwidth,angle=00]{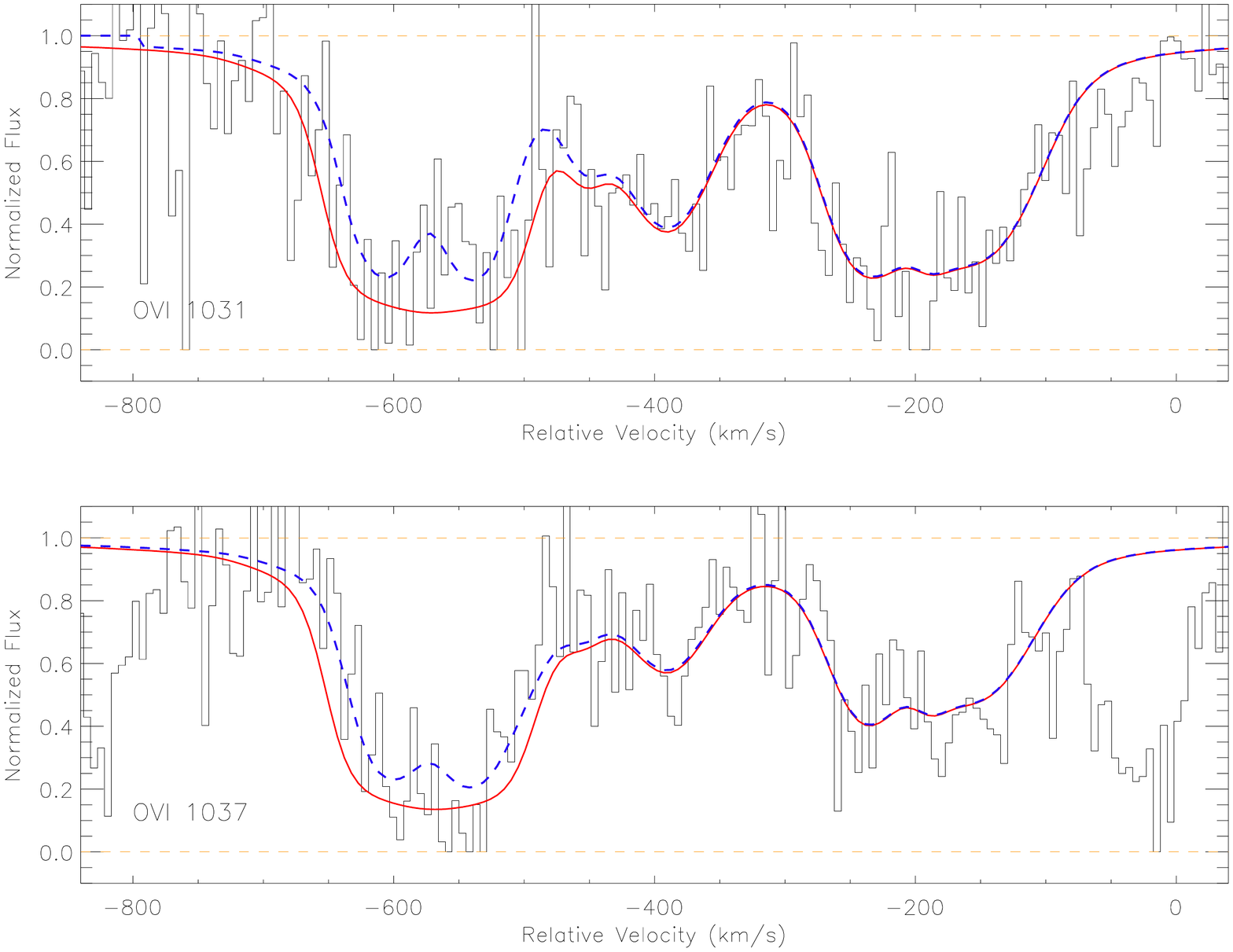}   
\caption{Synthetic spectrum for a model that adopts the CIE solutions of T11 for the two components at  $v = -613$~\kms\ and $-540$~\kms\ . The blue dashed curve is our adopted, photoionization model spectrum while the red solid curve is the same but with a CIE model with T  = 10$^{5.6}$ K, b$_{min}$(\OVI) = 20.3 \kms\, and $\log N(\OVI) \sim$~16.0 (ten times larger than our adopted column density) for the two components in question. The CIE model does not match the observed \OVI\ absorption. This demonstrates that although the \OVI\ profile is partially saturated, the adopted \OVI\ column density cannot be arbitrarily large to account for the CIE solution with $\log(T) \approx$ 5.6.  }    
\label{tripp_ovi}  
\end{figure*}

In the absence of information about \OVI\ absorption, T11 favored a collisional ionization origin for the \NeVIII\ and \NV\ absorption in systems A and B. In Fig.~\ref{fig_compare} we revisit their CIE models for the two System A components at $v = -613$ and $-540$ \kms ($v = -317$ and $-247$ \kms\ in T11, whose zero-point redshift is $z = 0.927$). Using the total hydrogen column density, $N_{\rm H}$, and metallicity, $\rm [X/H]$, as given in Table~S2 of T11, we calculate the absolute column densities of different high-ions as a function of gas temperature under CIE conditions. The temperatures, as derived by T11 from the \NeVIII\ to \NV\ column density ratios, are marked by the vertical dotted lines. It is apparent from the figure that the temperature solutions cannot reproduce the right amounts of absorption in \CIV\ and \OVI.  For both the components, the models predict $N(\OVI)\sim10^{16}$~cm$^{-2}$, which is $\sim10$ times higher than the measured values. The \OVI\ profiles in these components may suffer from saturation, causing us to underestimate their columns in our fits. However, we know the minimum b-value corresponding to the T11 model predicted temperature, and using that b-value, a model profile with $N(\OVI)\sim10^{16}$~cm$^{-2}$ exceeds the data (see Fig.~\ref{tripp_ovi}). Any non-thermal contribution to the \OVI\ Doppler parameter would further worsen the situation.  On the other hand, the models produce $\sim0.7$ dex lower $N(\CIV)$ than observed. The CIE solutions for the high ions based on \NV\ and \NeVIII\ are inconsistent with the observed column densities of \CIV\ and$/$or \OVI.

Our PI models in the previous section could explain \CIV, \NV, and \OVI\ column densities at these velocities arising from a single phase. The observed \NeVIII\ cannot be explained under photoionization equilibrium conditions, and we speculate that \NeVIII\ could be collisionally ionized in a stand alone phase. We note that there is a sweet spot of temperature (i.e. $\log (T/\rm K) = $~5.7--6.0) in which \NeVIII\ is the dominant species under both the CIE and non-CIE conditions. In this temperature range $N(\NeVIII)$ is higher than $N(\OVI)$ and $N(\MgX)$. For temperatures $>10^{6}$~K, the ion fraction of \NeVIII\ (\MgX) decreases (increases) sharply. As the \MgX\ is a non-detection with $\log N < 14.1$ (see Table~S2 of T11), a temperature of $> 10^{6}$~K is unlikely.

If we could reconcile the overproduction of \OVI\ in the collisionally ionized models of systems A and B that match \NV\ and \NeVIII, there would still be a problem with \CIV.  An additional photoionized phase would be needed to produce the \CIV\ absorption, but that phase would also have to produce intermediate ionization or other high ionization, which is already produced by the low ionization and higher ionization collisionally ionized phases. {\it} The model consistent with all of the data for systems A and B thus has two photoionized phases, and a $\log (T/\rm K) \sim 5.85$ collisionally ionized phase that produces the \NeVIII\ absorption.  System C could be similar, but it could instead have just one photoionized phase and a somewhat less hot collisionally ionized phase with $\log (T/\rm K) \sim 5.65$ in which the {\rm \OVI} and {\NeVIII} absorption arises.

% 
%==================================================================================
%==================================================================================
% 
%==================================================================================
%==================================================================================

\section{Galaxies Around the Quasar}   
\label{sec:galaxy}

As in Fig.~\ref{fig:HSTimage}, there are 4 galaxies detected near the quasar sightline, which we label G1--G4 following D03. Here we examine whether a single luminous galaxy or a group of galaxies is responsible for the complex absorption profile along this sightline.

One of the four galaxies, G2, was confirmed to be at $z =0.9289\pm0.0005$ based on the [\OII]$\lambda$3727 emission line at $7190\pm2$~\AA\ in a KPNO 4-m CryoCam spectrum (D03), which corresponds to an impact parameter of $68$~$\rm kpc$.  A higher $S/N$ spectrum of this same galaxy was analyzed by T11, who confirmed that it is close to the redshift of the absorber, and suggested that it has properties consistent with a post-starburst galaxy. Though T11 did not quote a precise redshift value, inspection of their Fig.~S2 gives a value consistent with the D03 value of $z_{\rm gal}=0.9289$.

An expanded view of galaxy G2 is shown in Fig.~\ref{fig:HSTimage}. The galaxy shows a bright nucleus and possibly two rings, suggesting that there has been a merger. The position angle of the bulge is $\Phi=14.1_{-4.1}^{+2.7}$ degree and that of the disk is $\Phi =88.5_{-6.6}^{+4.9}$ degree. Since the disk dominates the light (with bulge-to-total light fraction $B/T =0.39_{-0.03}^{+0.04}$), we determine that the quasar sight-line passes almost right along the projected minor axis of the disk. The inclination of the disk is moderate, measured as $41.5_{-6.3}^{+4.4}$ degree. With this inclination and a moderate opening angle, if absorption arises in an outflow along the minor axis it should be asymmetric in velocity due to the sightline passing through one side of the outflow but not through the other side. However if the opening angle is quite large it is also possible that the line of sight also grazes the opposite outflow cone, which would lead to highly redshifted absorption. This is a possible explanation for the origin of System C. The extended disk should also intercept the quasar line-of-sight so infalling gas could in principle produce absorption in this case as well.

We now consider the other three galaxies in the quasar field. Galaxy G1, at an impact parameter of 31~kpc, was spectroscopically identified in our Keck ESI spectrum, using the emission lines [\OIII]$\lambda$5008, H$\alpha$ and a sky-line blended [\NII]$\lambda$5685, shown in the right panel of Fig.~\ref{fig:HSTimage}. We used our own fitting program \citep[FITTER: see ][]{cwc-thesis} to compute best fit Gaussian amplitudes, line centers, and widths, in order to obtain emission line redshift. We determined the redshift of G1 to be $z=0.21441\pm0.00002$. A weak, low ionization metal-line absorber is known to be at a redshift of $z=0.21439$ (Muzahid et al., in preparation), consistent with this lower redshift. At this redshift, galaxy G1 is found to have a luminosity of $0.03L_B^*$. Galaxy G1 is clearly at a smaller redshift, and thus irrelevant to the present study.

Galaxy G3 was tentatively detected in a Fabry-Perot image tuned to [\OII]$\lambda$3727 at $z\sim0.93$ that was published in \cite{Thimm1995}, as mentioned by D03. At that redshift the impact parameter of G3 would be 74~$\rm kpc$, similar to that of G2. However, at $z=0.93$ G3 would be a $L_B = 0.44L*$ galaxy with a half-light radius of 5.5~$\rm kpc$.  This is implausibly large, thus G3 is more likely to be at a smaller redshift, despite the positive suggestion based on the Fabry-Perot image.

We have no information about the redshift of G4, which at $z=0.93$ would be a $L_B =0.12L*$ galaxy with a half-light radius of 2.6~$\rm kpc$ and an impact parameter of 113~$\rm kpc$.  It is possible that some metal-line absorption could arise along the quasar sight-line from G4, which would be within the halo of the $1.3L*$ galaxy, G2.

We conclude that one $1.3L*$ disk galaxy, G2 at an impact parameter of 68~$\rm kpc$, is definitely known to be at an appropriate redshift to produce the observed absorption. This galaxy shows signs that it has been influenced by a merger and has hints of AGN activity in its spectrum \citep[]{Tripp2011}. Another galaxy, G4, with a tenth the luminosity of G2 may also be at a similar redshift, but that cannot be confirmed without further observations. Lastly, we note that many dwarf galaxies could be embedded in the halo of galaxy G2, which would elude detection at this distance. We will return to discussion of this issue in Section~\ref{sec:discussion}.

%%================================================================================== 

\section{Discussion}
\label{sec:discussion}

The absorption complex at $z\sim0.92$ toward PG~1206+459 is remarkable in several respects. The absorption spans a velocity range of $\sim1400$~{\kms} though the bulk of the absorption is in systems A and B which range over $500$~\kms\ in velocity. The low-ionization and high-ionization transitions have similar, but not identical, kinematics. The strength of the \OVI\ and \NV\ absorption in this complex (i.e., $\log N({\OVI})=$~15.54$\pm$0.17 and $\log N({\NV}) =$~14.91$\pm$0.07) is the largest known for an intervening absorber, and the presence of \NeVIII\ absorption from all three systems is significant. The $\log N({\CIV}) \sim$~15.5 is also quite large. A partial Lyman break and spectral coverage of numerous Lyman series lines provide rigorous constraints on the metallicities of the various regions that produce the absorption, and most are constrained to have solar or super-solar metallicities for gas at an impact parameter of $68$~$\rm kpc$ from the nearest luminous galaxy. That spiral galaxy has a luminosity of $1.3L_B^*$ and a double-ring structure indicative of an interaction$/$merger.

\begin{figure*} 
\centerline{
\vbox{
\centerline{\hbox{ 
\includegraphics[width=0.75\textwidth,angle=270]{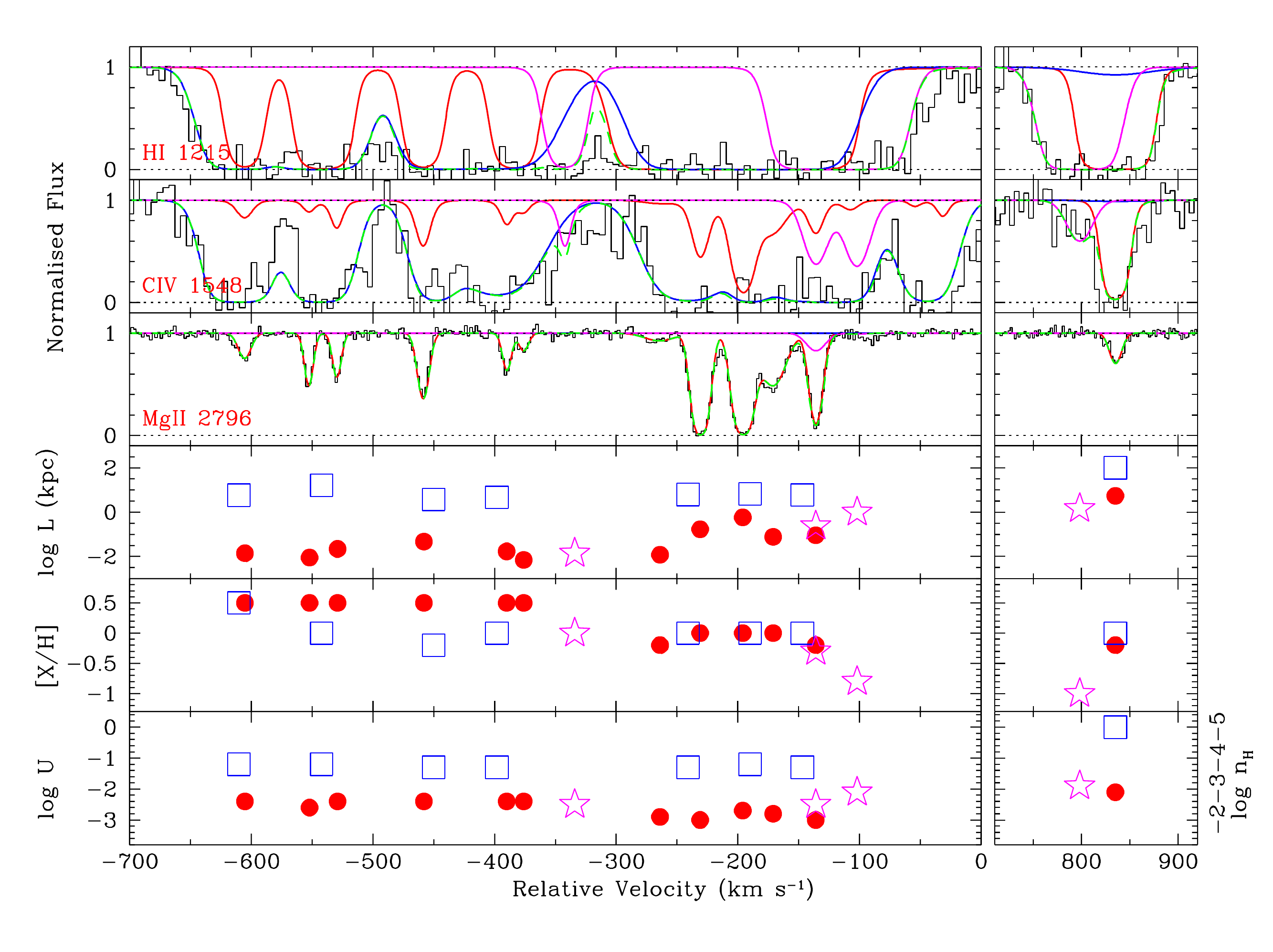}     
}}
}}
\vskip-0.5cm  
\caption{Summary of PI model parameters in individual low-/intermediate-/high-ionization components. In the top three panels, profiles and our adopted models of \lya, \CIV\lam1548, and \MgII\lam2796 are also shown. The color convention is the same as in Fig.~\ref{fig_modelABa} and Fig.~\ref{fig_modelCa}, red lines correspond to the low-ionization model, blue lines to the high-ionization model, magenta to intermediate-ionization and the \lya-only cloud at $\sim$-340 \kms. The total profile is shown in green. The bottom three panels show the sizes, metallicities, and ionization parameters for each model cloud, red circles for \MgII clouds, blue squares for \NV clouds, magenta stars for \SiIV clouds.}      
\label{fig:modelsum}  
\end{figure*}

Based on the new models presented in this paper, including constraints from our new COS spectrum, covering \OVI, we refine the constraints on the ionization parameters and metallicities of the different absorbing components. These constraints are summarized in Fig.~\ref{fig:modelsum}. For all three systems, A, B, and C, there are groups of clouds in two photoionized phases.  By a ``phase" we mean gas that falls within a certain range of density and temperature. The \MgII\ absorption is produced by photoionized ``clouds" with densities $-3 < \log n_{\rm H} < -2$~\cc and ionization parameters $-3 < \log U < -2$.  Across the system, there are 15 distinct low ionization ``clouds" like these, having line of sight thicknesses of tens to hundreds of $\rm pc$ and Doppler parameters of a few to $\sim$10~\kms. The higher ionization \CIV, \NV, and \OVI\ absorption arises in at least eight photoionized ``clouds" with densities $-4 < \log n_H < -3.7$~\cc and ionization parameters $-1.3 < \log U < -1$.  These lower density ``clouds" have thicknesses of several to 16~$\rm kpc$ and the absorption profiles are broader, with Doppler parameters of ten to twenty \kms. The \NeVIII\ absorption provides evidence for pervasive hotter gas, $\log T \sim 5.85$, spanning similar ranges of velocity.  The redshift of galaxy G2, corresponding to zero velocity in Fig.~\ref{fig:modelsum}, falling redward of all of the system A and B absorption, but 800~\kms\ blue-ward of System C. Keeping in mind a mental picture of the multiple phases of gas along this complex sightline, and their possible relationships, we revisit the question of the physical origin of this absorption line system.

%==================================================================================  
%%================================================================================== 

\subsection{Comparison to Previous Studies of This System}

As mentioned in the introduction, this same system was studied extensively in previous papers by our group (D03) and more recently by T11. We have revisited the system constraints because of the new wavelength coverage, particularly \OVI, afforded by our Cycle-19 COS data (program ID: 12466).

D03 applied similar component-by-component {\sc cloudy} modeling methodology as utilized in this study. We have used the ionizing radiation field (HM01), instead of HM96 \citep{Haardt1996} as used in D03. However, this only changes the constrained ionization parameters by a couple tenths of a dex since the normalization of the \HI\ ionizing radiation has changed from $\log n_{\rm \gamma} = -5.2$ to $-5.0$~\cc. This change certainly does not qualitatively change our conclusions. The physical conditions of the low ionization phase are confirmed by the present study, with metallicities of some of the individual components now more precisely constrained by the higher resolution COS coverage of the Lyman series lines from Cycle 17 (program ID: 11741) as also presented in T11.

D03 found that the higher ionization gas that gave rise to the \CIV\ and \NV\ absorption in the $HST/$STIS spectrum, if photoionized, could also produce the equivalent width of \OVI\ seen in the low resolution $HST/$FOS spectrum. That study predicted the appearance of higher resolution \OVI\ profiles (see Fig.~9 of D03). The \OVI\ observations presented in this paper do agree with photoionized models for the \CIV, \NV, and \OVI\ absorption. This does, however, leave the \NeVIII, which was not known at the time of the earlier study, unexplained and does call for a separate, collisionally ionized phase.

T11 used the ratio of \NeVIII\ to \NV\ absorption to constrain the temperature of the collisionally ionized gas responsible for the \NeVIII\ absorption as $\log T \sim 5.5$~K. In the present paper we have shown that a model with this temperature over-predicts \OVI\ by up to an order of magnitude, even when accounting for possible saturation, and also fails to produce \CIV\ absorption as observed by a factor of 5. We therefore favor separate gas phases, a photoionized one that produces \CIV, \NV, and \OVI, and a warmer, collisionally ionized one ($\log T \sim 5.85$) that produces the \NeVIII\ absorption. The present solution is somewhat more complex because it requires numerous low density ``clouds" along the line of sight which have higher density structures moving with them, all surrounded by a higher temperature diffuse medium. It is clear that availability of a very large number of different ionization states of different chemical transitions is essential to deriving the complete picture of an absorption line system. In this case adding the \CIV\ and \OVI\ was pivotal to deriving a more accurate physical model of the gas.

\subsection{Comparison to Other Absorption Systems}
The \OVI\ phase in z $\approx$ 0.93 absorber towards PG1206+459 was also included in the study of z$<$1 LLSs by \cite{Fox2013}. They studied \OVI\  in 23 systems that had been shown to be bimodal in metallicity \citep{Lehner2013}. The $N(\OVI)$ and $\Delta v_{90}(\OVI)$ (i.e. velocity spread) in their systems are both correlated with metallicity, leading to a natural interpretation as tracing outflows \citep{Fox2013}. 

The $z = 0.3985$ absorber towards Q~0122-003 \citep{Muzahid2015} has very strong \NV\ and \OVI\ (still factors of 2 less than our system), as well as a large velocity spread ($\sim$500 \kms, half as wide as ours). The sightline passes through the minor axis of $\sim$0.5~$L_B^*$ galaxy at a large impact parameter of 163~$\rm kpc$. The high ionization gas, traced by \CIV{},\NV{}, and \OVI{}, has super solar metallicity and is shown to be photoionized. These facts led \citet{Muzahid2015} to conclude that the high ionization absorbing clouds were produced by a powerful galactic outflow. It is also important to note that there are multiple distinct (separated in velocity) high ionization clouds along the line of sight in the \cite{Muzahid2015} absorber, each with a thickness $\sim$10~$\rm kpc$. The metallicity, temperature, ionization parameter, and thickness is thus very similar to the absorber we are studying here. {\NeVIII} is not covered in the spectrum of the \zabs~$=$~0.3985 absorber, however if it is detected then it would have to arise in a separate phase than the {\OVI}, just as in the PG~1206+459 absorber.

However, in contrast to our \zabs~$=$~0.93 absorber, the \zabs~$=$~0.3985 system has a lower ionization phase with a much smaller metallicity of $\log Z \sim -1.4$ than the high ionization phase. \cite{Muzahid2015} suggested that the lower metallicity, high density gas was infalling to the disk, similar to the interpretation given by \cite{Ribaudo2011} for the \zabs~$=$~0.274 toward PG~1630+377 absorber. We do not see such low metallicity gas in the present system, which has a total $\log N({\HI})<17$ as compared to $\log N({\HI})\sim$19 for the lower redshift absorber. That difference may simply be the result of whether the random line of sight passes through infalling gas or not. However another difference, the presence of System C in our \zabs~$=$0.93 absorber, leading to a much larger total velocity spread ($>$1000~{\kms}), may also be important to consider when discussing the origin of the absorber. For an outflow with an extremely large opening angle, it is possible that System C arises in the redshifted outflow cone, leading to its large (800~\kms) redshifted velocity. Alternatively, System C could just be produced by a related type of gas cloud which has a relatively small filling factor (due to its high relative velocity perhaps) but just happens to be along the line of sight to PG~1206+459. Its properties are similar to those of the population of weak \MgII\ absorbers \citep[]{Rigby2002,Narayanan2008} 

\cite{Meiring2013} studied three systems at \zabs~$\sim$0.7 which have \NeVIII\ and \OVI\ detected, as well as high metallicity lower ionization gas. However, with $\log N({\OVI}) \sim$14.4, these systems were simpler, and weaker, than the \zabs~$=$~0.93 absorption complex toward PG~1206+459. In fact, they are similar to system C taken alone, which has $\log N({\OVI})=$~14.45. \cite{Meiring2013} model the absorbers as multiphase structures with cool clouds with sizes $<4$~$\rm kpc$, which are unstable and expanding while moving through a diffuse, hotter surrounding medium. The interface between the cool cloud and hot medium is thought to produce the {\OVI} and {\NeVIII} absorption in these systems, with a temperature of about $\log T = 5.7$.  We found similar conditions for system C, also with {\OVI} and {\NeVIII} arising in a separate phase. Several other studies of low$/$intermediate redshift absorbers have also yielded the conclusion of {\OVI} and {\NeVIII} arising in collisionally ionized gas \citep[]{Savage2011,Narayanan2012}, sometimes citing an unrealistically large pathlength of $>$1~$\rm Mpc$ through the gas \citep[but see][]{Hussain2015,Hussain2017}.  

There are also some similarities between the conditions in the gas in our PG~1206+459 absorber and the high redshift ($2 < z < 3.5$) Lyman limit systems (LLSs) studied by \cite{Lehner2014}. Though most of the high redshift systems were DLAs or sub-DLAs there were several systems with $17 < \log N({\HI}) < 18$, and the average column density of {\OVI} for the 15/20 systems for which it was detected is $\log N({\OVI}) = 14.9$, just a few times smaller than the {\OVI} column density in the PG~1206+459 absorber. Typically, the {\NV} column density is an order of magnitude smaller than the {\OVI} column density, however, and the \cite{Lehner2014} systems span only 200--400~{\kms}, like system A or B alone. One system at \zabs~$=$2.18 towards Q~1217+499, does have $\log N({\OVI})$ and $\log N({\NV}) > 14.1$, but it is saturated over most of the profile. \cite{Lehner2014} interprets the {\OVI} in these high redshift LLSs as arising under non-equilibrium conditions in cooling gas, related to starburst galaxies. The latter conclusion is in part due to a measured correlation between the {\OVI} column density and the Doppler parameter of individual components. In some systems the component structure in {\CIV}, {\NV}, and {\OVI} is similar, like in our \zabs~$=$~0.93 absorber, but there are sometimes broader components in the {\OVI} as well. Perhaps the latter broad {\OVI} is produced by a structure similar to that which gives rise to {\NeVIII} in the case of PG~1206+459. But it also seems clear that low density photoionized gas gives rise to some of the {\OVI} absorption at high redshift as well.    

Much of the information available about \OVI\ in the CGM of galaxies in recent times has come from the COS Halos study of \cite{Tumlinson2011}. That study focused on low redshift ($0.1 < z < 0.36$), $\sim$$L*$  galaxies at impact parameters $<$150~$\rm kpc$ from quasar sightlines. \cite{Tumlinson2011} found that \OVI\ absorption with $\log N({\OVI})>14.2$ is detected within this impact parameter from the CGM of star-forming galaxies, with specific star formation rate $>10^{-11}$yr$^{-1}$ but not from passive galaxies with smaller specific star formation rates. Our higher redshift system has \OVI\ absorption, $\log N({\OVI})=$15.6 several times stronger than any of the galaxies in the COS Halos Survey.

%==================================================================================  
\begin{figure*} 
\begin{centering}
\includegraphics[width=\textwidth,angle=0]{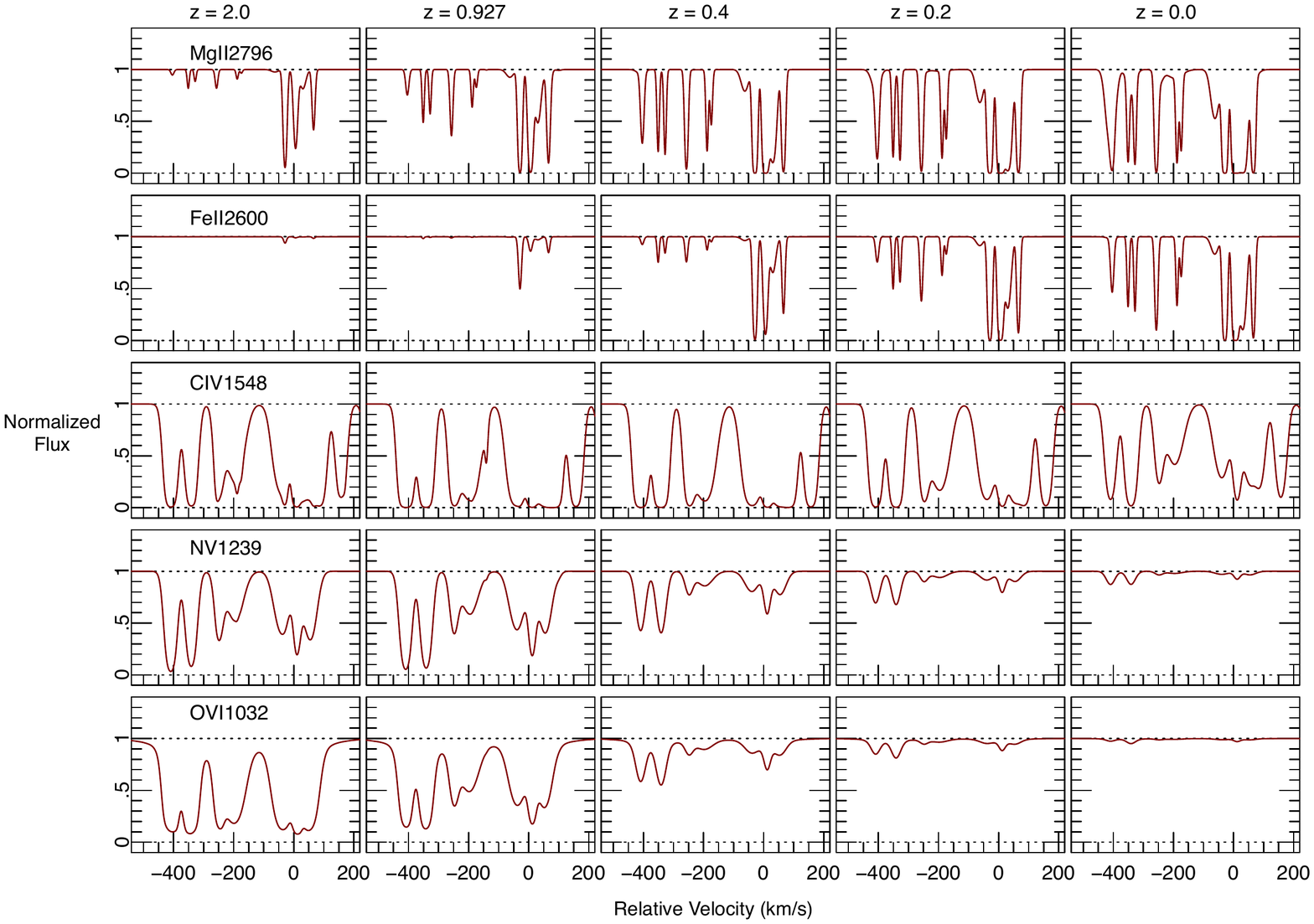}  
\vskip-1.5cm  
\caption{Redshift evolution of different absorption lines detected in the present absorber. An absorber with the same density, temperature, and metallicity as the one in our system, placed in the EBR at different redshifts, would be observed to look like the plotted profiles. The ions are indicated in the left-most columns. The redshifts are indicated in the top-most panels. The changes in profiles are due to the changing ionizing EBR at different redshifts. The physical processes/structures that produced the strongest intervening \OVI\ absorber at $z\sim1$ wouldn't have produced any detectable \OVI\ at $z\lesssim0.2$.}       
\label{fig:evolution}
\end{centering}
\end{figure*} 
%%================================================================================== 

\subsection{Evolution of the Observational Signatures of Absorption}

In interpreting the origins of a particular class of absorber, it can be very important to recognize that the evolving extragalactic background radiation alters the absorption signature of a given type of structure.  Our solutions from modeling the \zabs~$=$~0.93 absorber yielded a number of clouds with given densities, sizes, and metallicities. The philosophy behind the present thought experiment is the idea that certain processes (like outflows or inflows) or types of structures (such as tidal debris or high velocity clouds) that exist at \zabs~$=$~0.93 will also exist at other redshifts. We are simply considering what these types of processes or structures will have as their absorption signatures at different redshifts due to there being a different extragalactic background radiation field.

At a different redshift the very same clouds that produce the observed amount of absorption in {\MgII}, {\CIV}, {\OVI}, etc., will produce a different amount of absorption because the incident radiation field is more or less intense. To better understand this in the present case, we took the cloud properties in Tables~\ref{tab:PImodel} and ran {\sc cloudy} models with the HM01 extragalactic background field at redshifts $z=2$, $0.4$, $0.2$, and $0$. The simulated model profiles for the key constraining transitions {\MgII}, {\FeII}, {\CIV}, {\NV}, and {\OVI} are shown at the different redshifts, including the actual redshift of our system, in Fig.~\ref{fig:evolution}. Also, while the {\HI} column density for our system at \zabs~$=$~0.93 is $\log N({\HI}) =$~16.9, we find that a full Lyman limit break will occur at lower redshifts ($\log N({\HI}) = 18.2$ at $z=0$ and $17.9$ at $z=0.2$). Conversely, at $z=2$ we find that $\log N({\HI})$ is reduced to $16.6$.

From Fig.~\ref{fig:evolution} we can see that this, the structure producing the strongest known {\OVI} absorber, would not even have detected {\OVI} absorption at $z=0$, and at $z=0.2$ the {\OVI} absorption would be quite weak, not even strong enough to have been detected in the COS Halos survey. The same applies to the {\NV} absorption. These structures, with a scale of a few to 16~$\rm kpc$ and densities of $n_{\rm H} \sim$$10^{-3.8}$~{\cc}, would not be {\OVI} absorbers at low redshift. Even if those same structures have lower densities, so that the ionization parameter was the same at low redshift as for the PG~1206+459 absorber, structures of those same sizes would not have a large enough path-length for the lower densities in order to produce the strong {\OVI} and {\NV}. Thus {\OVI} absorbers of this type are present at $z\ge0.4$, but not at lower redshifts.

The strong {\OVI} absorbers that do exist at lower redshift must then either be of lower density and/or larger structure, the sizes of entire halos or intra-group medium, and photoionized \citep[e.g.,][]{Muzahid2014}, or must be hotter such that they are collisionally ionized \citep[e.g.,][]{Savage2010,Savage2011}. Similarly, Fig.~\ref{fig:evolution} shows that at $z=2$ the high ionization absorption from these types of structures would be significantly stronger than it is at \zabs~$=$~0.93. The {\MgII} absorption would be weaker, and the {\FeII} absorption would become negligible.

Let us also consider the evolving origin of {\NeVIII} absorption.  At $z=0.93$ in the PG~1206+459 absorber, at least for systems A and B, a $\log T \sim 5.85$ collisionally ionized phase could account for the observed {\NeVIII} absorption.  This could also happen at lower redshift, for gas of the same temperature.  Lower temperature gas could also give rise to {\OVI} in the same collisionally ionized phase with the {\NeVIII}, thus this is a possible origin of low redshift {\OVI} absorption.

The point of this thought experiment is to emphasize that whatever the physical origin of the \zabs~$=$~0.93 absorption complex, the same type of object at low redshift will not have the same absorption signature. In particular, it would not even give rise to {\OVI} absorption. Thus at low redshift ($z\lesssim0.2$), the population of {\OVI} absorbers must have a different origin.

\subsection{Physical Origin of the Absorber}

In order to see the unprecedented strong {\OVI}, {\NV}, and {\NeVIII} evident in the \zabs~$=$0.93 PG~1206+459 absorption complex the conditions have to be optimal in several ways. The $>$1000~{\kms} velocity spread is also extreme as compared to other absorbers, and it applies to both the low and high ionization gas. The multiphase structure of the absorption also provides important constraints on a realistic physical picture. The dozen or so lowest ionization phase clouds have densities of 0.003--0.01~\cc, while the high ionization clouds have densities an order of magnitude lower. The cloud extent along the line-of-sight is of order 10~$\rm kpc$ for the high ionization clouds, with the lower ionization clouds more than two orders of magnitude thinner. At impact parameter of $68$~$\rm kpc$, fitting in seven large clouds along the line-of-sight implies a relatively large filling factor. The kinematics of the low and high ionization clouds appear to be related implying that the low ionization gas is embedded in or adjacent to the high ionization gas. The \NeVIII\ absorption that is detected in systems A and B is likely to arise in a hotter ($\log T = 5.85$~K) collisionally ionized phase. Almost all the clouds are constrained to have high metallicities, solar or a few times the solar value. Given the relatively high metallicities of all the systems, we have also discussed the possibility that the blueshifted system A/B absorption and the redshifted system C absorption come from opposite cones in an outflow with a large opening angle.

Using an $HST$ image of the quasar field, we have learned that the structure of the nearest galaxy to the sightline (at an impact parameter of 68~$\rm kpc$) suggests a recent merger (see Fig.~\ref{fig:HSTimage}). Systems A and B have similar properties, but their low ionization gas is kinematically distinct, and system C is kinematically separated from system B by $\sim$800~{\kms}. D03 suggested an association of each of the systems, A, B, and C, with a different galaxy. Here we have ruled out two of the three additional candidate galaxies that could be responsible (galaxies G1 and G3 in Fig.~\ref{fig:HSTimage}), though G4, which is a tenth the luminosity of the confirmed galaxy, G2, could be related.  However, additional fainter galaxies, and galaxies at somewhat larger impact parameters could also be members of the same group as galaxy G2.

The absorber at  \zabs~$=$0.207 along the HE~0226-4110 line-of-sight \citep{Savage2011} may provide some hints about the relationship between absorption systems, galaxies, and groups. In that case, three galaxies are found within 300~{\kms} and 300~$\rm kpc$ of the absorber, two with luminosity 0.25L$_*$ and the other with luminosity 0.05$L_*$ \citep[see also][]{Mulchaey2009}. The  lower ionization transitions in that absorber, such as {\CIII}, {\OIII}, and {\OIV}, were found to arise in a photoionized phase with cloud size of 57~$\rm kpc$, but the {\OVI} and {\NeVIII} cannot be photoionized because the low densities needed would require unrealistically large cloud sizes.  Though \cite{Savage2011} suggested an origin of the {\OVI} and {\NeVIII} in collisionally ionized gas with $\log T \sim 5.73$, \cite{Mulchaey2009} favor its origin in conductive fronts at the boundary between the low ionization clouds and a much hotter halo gas or intra-group medium. Since our \zabs~$=$0.93 system is at higher redshift, and subject to a more intense EBR, the same photoionized clouds as gave rise to {\CIII}, {\OIII}, and {\OIV} at \zabs~$=$0.207 would now produce significant {\CIV}, {\NV}, and {\OVI} absorption. It could be that our {\NeVIII} (for which we derived $\log T =5.85$~K for collisional ionization) would arise in a conductive interface layer. Again a hotter ($\log T> 6$) region would surround these clouds, and possibly confine them. More directly, the Milky Way Galaxy has been found to have a hot halo with temperature $\log T = 6.1$ to $6.4$, and with a density of $\sim 2\times 10^{-5}$~{\cc} at a distance of $\sim 100$~$\rm kpc$, derived by combining X-ray absorption and emission measures \citep[]{Gupta2012}. If such halos extend to $50$--$100$~$\rm kpc$ around many other galaxies, as one would expect since the Milky Way is not unique, then this hot medium may affect the properties of clouds of gas around a variety of galaxies, both through its potential to confine, and the potential of interaction of moving clouds with their hotter surroundings.

Although hot halo gas and/or an intra-group medium may be important for producing the observed phase structure in our \zabs~$=$0.93 absorber toward PG~1206+459, it does seem that there is also compelling evidence for an outflow. In fact the hotter confining gas could be produced by an outflow as well. Whether a starburst outflow is the only mechanism responsible for the PG~1206+459 \zabs~$=$0.93 absorption complex or not, it seems almost certain that it is a factor. The high metallicities at large distances from the closest galaxy are one indication. The large velocity spread of the absorption is another. Although 1000~{\kms} is a large velocity, even for a strong outflow, it is not uncommon. \cite{Sell2014} studied a sample of twelve $0.45 < z < 0.7$ galaxies with $>1000$~{\kms} outflows and found that the majority had direct evidence for recent mergers, usually tidal debris.  They were able to show that the winds, which were observed through the resolved kinematics of interstellar {\MgII} absorption from the galaxies, are driven by star formation in a compact core region, and not by AGN activity. Given the orientation of the PG~1206+459 absorbing galaxy and its apparent merger activity, evidence is building that indeed a starburst outflow is producing most of the absorbing gas. Although it would require a large opening angle, it is possible that systems A, B and system C arise from opposite outflow cones. There is even evidence for molecular gas at distances of $\sim 10$~$\rm kpc$, moving at speeds up to $1000$~{\kms} around a $z=0.7$ starburst galaxy, studied by \cite{Geach2014}. The kind of phase structure that we observe in the PG~1206+459 is quite plausible in such an event. 

%%================================================================================== 

\section{Conclusions}
\label{sec:conclusions}

We present a detailed analysis of a partial Lyman limit system with $\log N(\HI)\sim17.0$ at \zabs~$=0.93$ detected towards the quasar PG~1206+459. The absorber was studied previously by D03 and T11. Here we present a medium resolution NUV spectrum obtained with the COS G185 grating that covers the \OVI\ doublet from the absorber and an ACS image of the quasar field. We measured a total $N(\OVI)$ of $10^{15.54\pm0.17}$~\sqcm, which is the highest \OVI\ column density ever measured in any intervening system. The absorber also shows the highest velocity spread of $>1000$~\kms\ which we separated into system A ($-650 < v < -350$~\kms), system B ($-300 < v < -50$~\kms), and system C ($+750 < v < +900$~\kms) following D03.  
%%%%

Consistent with previous studies, we have found that all three systems (A, B, and C) show a multiphase structure. While the densities of the high-ionization phases ($\sim 0.06\rm cm^{-3}$) are about an order of magnitude lower than the low-ionization phase ($\sim 0.003~\rm cm^{-3}$), both phases show near-solar to super-solar metallicities. The intermediate-ionization phase required to explain the \SiIV\ absorption in system B, however, shows somewhat lower metallicities ($\log Z \sim -0.3$ to $-0.8$).  The photoionization solutions for the high-ionization gas phases can explain all of the \CIV, \NV, and \OVI\ absorption. Therefore, the \NeVIII\ absorption, as reported in T11, must stem from a separate, seemingly collisionally ionized gas phase with temperatures of $T\sim10^{5.85}~\rm K$. The CIE/non-CIE temperature solutions derived by T11, assuming \NV\ and \NeVIII\ are in the same gas phase,   the column densities of \OVI\ that we measured from the new COS G185 grating observations and significantly under-produce the \CIV\ column densities constrained from previous STIS observations.                         
%%%          

Analyzing the ACS image of the quasar field, we found that the sightline passes through the projected minor-axis of the known luminous  ($1.3L_*$), nearby ($68~\rm kpc$), host-galaxy at $z_{\rm gal}=0.9289$. A ring-like structure seen in the image suggests recent merger events in the host-galaxy. T11 classified the galaxy as post-starburst. Therefore, it is most likely that the bulk of the absorbing gas arise from an outflow from the host-galaxy as was also suggested by T11. The kinematics are consistent with systems A, B, and C arising from opposite outflow cones. However, the presence of occasional, relatively lower metallicity absorption components possibly suggests that many other faint galaxies or processes other than just outflow may be contributing to the absorption complex. Complete information about faint continuum/line emitting galaxies around the quasar using future integral field spectrograph observations is indispensable for further insights about this spectacular absorber.             
%%%% 

Finally, we demonstrate how the evolving EBR substantially alters the strengths of different absorption lines. For example, the strongest \OVI\ absorber that we studied here would not have been detected in COS-Halos survey at $z\lesssim0.2$. We thus concluded that any strong \OVI\ absorbers that exist at lower redshift must have a different origin.

\noindent  ~\\ 
{\bf{ACKNOWLEDGEMENTS:}}~  We thank the referee for a helpful and detailed report that improved this work. Support for this research was provided by NASA through grant HST GO-12466 from the Space Telescope Science Institute, which is operated by the Association of Universities for Research in Astronomy, Inc., under NASA contract NAS 5-26555. G.G.K acknowledges the support of the Australian Research Council through the award of a Future Fellowship (FT140100933). Some of the data presented here were obtained at the W. M. Keck Observatory, which is operated as a scientific partnership among the California Institute of Technology, the University of California, and the National Aeronautics and Space Administration.  The Observatory was made possible by the generous financial support of the W. M. Keck  Foundation. Observations  were  supported  by Swinburne Keck program {\bf 2014A\_W178E}.

%================== DO not delete ============  
%---------------------------------------------
\def\aj{AJ}%
\def\actaa{Acta Astron.}%
\def\araa{ARA\&A}%
\def\apj{ApJ}%
\def\apjl{ApJ}%
\def\apjs{ApJS}%
\def\ao{Appl.~Opt.}%
\def\apss{Ap\&SS}%
\def\aap{A\&A}%
\def\aapr{A\&A~Rev.}%
\def\aaps{A\&AS}%
\def\azh{AZh}%
\def\baas{BAAS}%
\def\bac{Bull. astr. Inst. Czechosl.}%
\def\caa{Chinese Astron. Astrophys.}%
\def\cjaa{Chinese J. Astron. Astrophys.}%
\def\icarus{Icarus}%
\def\jcap{J. Cosmology Astropart. Phys.}%
\def\jrasc{JRASC}%
\def\mnras{MNRAS}%
\def\memras{MmRAS}%
\def\na{New A}%
\def\nar{New A Rev.}%
\def\pasa{PASA}%
\def\pra{Phys.~Rev.~A}%
\def\prb{Phys.~Rev.~B}%
\def\prc{Phys.~Rev.~C}%
\def\prd{Phys.~Rev.~D}%
\def\pre{Phys.~Rev.~E}%
\def\prl{Phys.~Rev.~Lett.}%
\def\pasp{PASP}%
\def\pasj{PASJ}%
\def\qjras{QJRAS}%
\def\rmxaa{Rev. Mexicana Astron. Astrofis.}%
\def\skytel{S\&T}%
\def\solphys{Sol.~Phys.}%
\def\sovast{Soviet~Ast.}%
\def\ssr{Space~Sci.~Rev.}%
\def\zap{ZAp}%
\def\nat{Nature}%
\def\iaucirc{IAU~Circ.}%
\def\aplett{Astrophys.~Lett.}%
\def\apspr{Astrophys.~Space~Phys.~Res.}%
\def\bain{Bull.~Astron.~Inst.~Netherlands}%
\def\fcp{Fund.~Cosmic~Phys.}%
\def\gca{Geochim.~Cosmochim.~Acta}%
\def\grl{Geophys.~Res.~Lett.}%
\def\jcp{J.~Chem.~Phys.}%
\def\jgr{J.~Geophys.~Res.}%
\def\jqsrt{J.~Quant.~Spec.~Radiat.~Transf.}%
\def\memsai{Mem.~Soc.~Astron.~Italiana}%
\def\nphysa{Nucl.~Phys.~A}%
\def\physrep{Phys.~Rep.}%
\def\physscr{Phys.~Scr}%
\def\planss{Planet.~Space~Sci.}%
\def\procspie{Proc.~SPIE}%
\let\astap=\aap
\let\apjlett=\apjl
\let\apjsupp=\apjs
\let\applopt=\ao
%---------------------------------------------
\bibliographystyle{mn}
\bibliography{ben}
%---------------------------------------------

\end{document}